\documentclass[12pt]{article}
\pdfoutput=1 

\usepackage{amsmath}
\usepackage{graphicx,tabularx}
\usepackage{url}
\usepackage{footmisc}
\usepackage{amsfonts}
\usepackage{cancel}
\usepackage{color}
\usepackage{multirow} 
\usepackage{amssymb}
\usepackage{pifont}
\usepackage{subfigure}
\usepackage{epstopdf}
\usepackage{slashed}
\usepackage{comment}
\usepackage{booktabs} 
\usepackage{extarrows}
\usepackage{color}
\usepackage{feynmp}
\IfFileExists{latexsym.sty}{\usepackage{latexsym}}{}
\IfFileExists{mathtools.sty}{\usepackage{mathtools}}{}
\allowdisplaybreaks[4]
\usepackage{jheppub} 

\def\Tr{\text{Tr}}

\title{Searching for the $W \gamma$ decay of a charged Higgs boson}
\author[a]{Heather E.\ Logan,}
\author[a]{Yongcheng Wu}
\affiliation[a]{Ottawa-Carleton Institute for Physics, Carleton University, 1125 Colonel By Drive, Ottawa, Ontario K1S 5B6, Canada }
\emailAdd{logan@physics.carleton.ca}
\emailAdd{ycwu@physics.carleton.ca}

\abstract{
We study the prospects for charged Higgs boson searches in the $W \gamma$ decay channel.  This loop-induced decay channel can be important if the charged Higgs is fermiophobic, particularly when its mass is below the $WZ$ threshold.  We identify useful kinematic observables and evaluate the future Large Hadron Collider sensitivity to this channel using the custodial-fiveplet charged Higgs in the Georgi-Machacek model as a fermiophobic benchmark.  We show that the LHC with 300~fb$^{-1}$ of data at 14~TeV will be able to exclude charged Higgs masses below about 130~GeV for almost any value of the SU(2)$_L$-triplet vacuum expectation value in the model, and masses up to 200~GeV and beyond when the triplet vacuum expectation value is very small.  We describe the signal simulation tools created for this analysis, which have been made publicly available.
} 

\begin{document}

\titlepage

\maketitle

\newpage


\flushbottom

\section{Introduction}

The discovery of the Higgs boson at the CERN Large Hadron Collider (LHC)~\cite{Aad:2012tfa,Chatrchyan:2012xdj} represents the first experimental evidence for a (possibly) fundamental scalar particle.  This naturally raises the question of whether there are more fundamental scalars; in particular, whether the Higgs sector is the minimal one predicted in the Standard Model (SM) or whether there are additional Higgs bosons.

Most extensions of the SM Higgs sector contain electrically-charged Higgs bosons $H^{\pm}$, which require very different experimental search strategies than do neutral Higgs bosons.  The standard charged Higgs searches at the LHC exploit the charged Higgs couplings to SM fermion pairs, which are expected in models in which the charged Higgs comes from an additional SU(2)$_L$ doublet of scalars.  These searches comprise charged Higgs production in top quark decays with the charged Higgs decaying to $\tau\nu$~\cite{Aad:2014kga,Khachatryan:2015qxa}, $c \bar s$~\cite{Aad:2013hla,Khachatryan:2015uua}, or $c \bar b$~\cite{Sirunyan:2018dvm}, as well as associated production of a charged Higgs and a top quark with the charged Higgs decaying to $\tau\nu$~\cite{Aaboud:2016dig,Khachatryan:2015qxa} or $t \bar b$~\cite{Aad:2015typ,Khachatryan:2015qxa}.  Searches for a charged Higgs produced in the decay of a heavier neutral Higgs have also been proposed for the LHC~\cite{deFlorian:2016spz}.

Fermiophobic charged Higgs bosons appear in a number of models including the Georgi-Machacek (GM) model~\cite{Georgi:1985nv,Chanowitz:1985ug}, the Stealth Doublet model~\cite{Enberg:2013ara,Enberg:2013jba}, and certain parameter regions of the Aligned two-Higgs-doublet model (2HDM)~\cite{Pich:2009sp}.
The fermiophobic charged Higgs in the GM model, denoted $H_5^{\pm}$ because it is a member of a fiveplet of the custodial symmetry, couples at tree level to $W^{\pm}Z$ with strength proportional to the vacuum expectation value (vev) of the SU(2)$_L$ triplets in the model.  Dedicated searches have been performed at the LHC for $H_5^{\pm}$ produced in vector boson fusion and decaying to $W^{\pm}Z$~\cite{Sirunyan:2017sbn,Aaboud:2018ohp,CMS:2018ysc}; these have focused on charged Higgs masses above 200~GeV.  A fermiophobic charged Higgs can also decay into $W \phi$ (where $\phi$ is a neutral scalar) and, at one loop, into $W \gamma$.  A dedicated search for $H^{\pm} \to W^{\pm} h$, where $h$ is the 125~GeV SM-like Higgs boson, has been performed by ATLAS~\cite{Aad:2013dza} in the context of a cascade decay $H^0 \to H^{\pm} W^{\mp} \to h(\to b \bar b) W^{\pm} W^{\mp}$ in a two-Higgs-doublet model.  Several searches for $W^{\pm} h$ resonances have also been made at the LHC~\cite{Aad:2015yza,Sirunyan:2017wto,Aaboud:2017ahz,Aaboud:2017cxo,Sirunyan:2018qob} for resonance masses as low as 300~GeV; these have been interpreted in the context of a spin-1 resonance, but could be recast for a charged scalar.  Searches for a $W^{\pm}\gamma$ resonance have been performed at the LHC~\cite{Aad:2014fha,Aaboud:2018fgi} for resonance masses as low as 275~GeV, again in the context of a narrow spin-1 resonance.  None of these LHC searches to date have considered resonance masses below 200~GeV.

In this paper we study the prospects for light charged Higgs boson searches in the decay channel $H^{\pm} \to W^{\pm} \gamma$.  This decay first appears at one loop\footnote{An exception is the charged Higgs arising from an isospin singlet with nonzero hypercharge, for which the decay to $W \gamma$ is forbidden at one-loop level~\cite{Cao:2017ffm}.}~\cite{Arhrib:2006wd,Enberg:2013jba,Ilisie:2014hea,Degrande:2017naf}, and hence its branching ratio is typically very small if tree-level decays to fermion pairs or $W^{\pm}Z$ are available.  However, for a fermiophobic charged Higgs with mass below the $W^{\pm}Z$ threshold, the branching ratio into $W^{\pm}\gamma$ can dominate~\cite{Enberg:2013jba,Ilisie:2014hea,Degrande:2017naf}, especially if the coupling to $W^{\pm}Z$ is suppressed due to a small triplet vev in the GM model or induced only at one loop as in the Stealth Doublet model and the Aligned 2HDM.  We will therefore focus on charged Higgs masses below 200~GeV.

This paper is organized as follows.  In Sec.~\ref{sec:distributions} we examine the general form of the loop-induced $H^{\pm} W^{\mp} \gamma$ vertex and derive the key kinematic distribution that we will use to discriminate the charged Higgs decay from backgrounds.  We also discuss the possible contributions to the loop-induced effective couplings that control this distribution.  In Sec.~\ref{sec:benchmark} we choose the fermiophobic $H_5^{\pm}$ in the GM model as a concrete benchmark.  After a brief description of the model to set our notation, we summarize the relevant decay modes and discuss the most important charged Higgs production processes in the low-$H_5^{\pm}$-mass region.  We focus on Drell-Yan production of $H_5^{\pm}$ in association with another member of the scalar custodial fiveplet because of its large cross section even in the small triplet vev limit and its independence from the choice of model parameters.  

In Sec.~\ref{sec:search} we perform a sensitivity study for the $H_5^{\pm} \to W^{\pm} \gamma$ channel and evaluate the exclusion reach for 300~fb$^{-1}$ at the 14~TeV LHC.  We describe our implementation of the loop-induced decays via effective couplings in a new Universal FeynRules Output (UFO)~\cite{Degrande:2011ua} model file to be used with version 1.4.0 of the model calculator GMCALC~\cite{Hartling:2014xma} (these have been made publicly available).  We simulate the dominant backgrounds and give an optimized set of cuts.  Our main result is a projection for the 95\% confidence level upper limit on the signal fiducial cross section as a function of the charged Higgs mass, which we then interpret as an upper limit on BR($H_5^{\pm} \to W^{\pm}\gamma$) and an exclusion reach in the GM model parameter space.  In particular, we find that the LHC with 300~fb$^{-1}$ of data at 14~TeV will be able to exclude $H_5^{\pm}$ masses below about 130~GeV for almost any value of the triplet vev, and masses up to 200~GeV and beyond when the triplet vev is very small.  Finally in Sec.~\ref{sec:conclusions} we summarize our conclusions.  Details of our choice of the parameter benchmark in the GM model and the form factors in the limit of small triplet vev are given in Appendices~\ref{app:LSH} and \ref{app:FormFactor}, respectively.

\section{$H^+ \to W^+ \gamma$ decay}
\label{sec:distributions}

The decay amplitude for $H^+(k+q) \to W^+_{\nu}(k) \gamma_{\mu}(q)$ is forced by electromagnetic gauge invariance to take the form~\cite{Ilisie:2014hea}
\begin{equation}
	\mathcal{M} = \Gamma^{\mu\nu} \varepsilon^{W*}_\nu(k) \varepsilon^{\gamma*}_\mu(q), 
	\quad {\rm with} \quad 
	\Gamma^{\mu\nu} = (g^{\mu\nu}k\cdot q-k^\mu q^\nu)S + i\epsilon^{\mu\nu\alpha\beta}k_\alpha q_\beta \tilde{S},
	\label{eq:vertex}
\end{equation}
where $k$ and $q$ are the four-momenta and $\varepsilon^W_{\nu}(k)$ and $\varepsilon^{\gamma}_{\mu}(q)$ are the polarization vectors of the $W$ boson and the photon, respectively.  

The form factors $S$ and $\tilde S$ for $H^+ \to W^+ \gamma$ have been computed in 2HDMs in Refs.~\cite{Arhrib:2006wd,Enberg:2013jba,Ilisie:2014hea} (Ref.~\cite{Arhrib:2006wd} also considered the Minimal Supersymmetric Standard Model (MSSM)) and in the GM model in Ref.~\cite{Degrande:2017naf}.  In a CP-conserving theory, the scalar form factor $S$ receives contributions from loops of fermions, scalars, and gauge bosons, while the pseudoscalar form factor $\tilde S$ receives contributions only from loops of fermions; this implies that for a fermiophobic charged scalar, $\tilde S \to 0$.  Furthermore, while $S$ and $\tilde S$ are complex in general, their imaginary parts arise only if a contributing loop diagram can be cut yielding an on-shell tree-level two-body decay.  While we maintain full generality in this section, it will be useful to keep in mind the fact that the $H^+ \to W^+ \gamma$ decay is most interesting phenomenologically when competing decays to on-shell two-body final states and to fermion pairs are absent, i.e., when both form factors are real and $\tilde S \to 0$.

The vertex in Eq.~(\ref{eq:vertex}) leads to the $H^+ \to W^+ \gamma$ decay partial width
\begin{equation}
	\Gamma(H^+ \to W^+ \gamma) = \frac{m_{H^+}^3}{32 \pi} 
	\left[1 - \frac{m_W^2}{m_{H^+}^2} \right]^3 \left[ |S|^2 + |\tilde S|^2 \right],
\end{equation}
where $m_{H^+}$ is the mass of $H^+$ and $m_W$ is the mass of the $W$ boson.

\subsection{Differential distributions}
\label{sec:differential}

In practice, the $W$ boson will be reconstructed from its decay products, providing an additional experimental handle on the structure of the $H^+ W^- \gamma$ vertex via the $W$ polarization.  Allowing the $W$ boson to decay leptonically to $\ell^+ \nu$, the square of the matrix element takes the form
\begin{eqnarray}
	|\mathcal{M}|^2 &\propto& \Gamma^{\mu\nu}\Gamma^{\rho\sigma *}\varepsilon^{\gamma*}_\mu\varepsilon^{\gamma}_\rho\textbf{ Tr}(\slashed{p}_\nu\gamma_\sigma P_L \slashed{p}_\ell \gamma_\nu) \nonumber \\
	&=& \frac{m_W^2}{2} \left\{ 8 (p_\ell\cdot q)^2 \left[ |S|^2+|\tilde{S}|^2 \right] 
- 4 (p_\ell\cdot q) (m_{H^+}^2-m_W^2) \left[ |S|^2+|\tilde{S}|^2-2\textbf{Re}(S\tilde{S}^*) \right] \right. \nonumber \\
	& & \left.+ (m_{H^+}^2-m_W^2)^2 \left[ |S|^2+|\tilde{S}|^2-2\textbf{Re}(S\tilde{S}^*) \right] \right\},
	\label{eq:Msq}
\end{eqnarray}
where $p_{\ell}$, $p_{\nu}$, and $q$ are the four-momenta of the final-state lepton $\ell^+$, neutrino, and photon, respectively.  Here we have assumed that the $W$ boson is emitted on shell and the $W$ propagator dependence in $|\mathcal{M}|^2$ is omitted, which for an on-shell $W$ is just an overall multiplicative factor.  We have also neglected the final-state fermion masses.

In particular, the square of the matrix element can be expressed as a quadratic polynomial in the experimentally-observable kinematic invariant $p_{\ell} \cdot q \equiv p_{\ell}^{\mu} q_{\mu}$, the kinematically-accessible range of which is $[0, (m_{H^+}^2-m_W^2)/2]$.
It is convenient to reparameterize the form factor and momentum dependence of the kinematic distribution in Eq.~(\ref{eq:Msq}) in terms of the ratios
\begin{equation}
	r \equiv \frac{\tilde S}{S}, \qquad
	K \equiv \frac{p_{\ell} \cdot q}{(m_{H^+}^2 - m_W^2)/2} \in [0,1],
\end{equation}
where a fermiophobic charged Higgs corresponds to $r \to 0$.  The kinematic distribution in Eq.~(\ref{eq:Msq}) can then be rewritten as
\begin{eqnarray}
	|\mathcal{M}|^2 \propto 2 K^2 \left[ 1 + |r|^2 \right]
	+ (- 2 K + 1) \left[1 + |r|^2 - 2 \textbf{Re}(r) \right].
	\label{eq:Msqr}
\end{eqnarray}
This function is a parabola in $K$ with its minimum at 
\begin{equation}
	K_{\rm min} = \frac{1 + |r|^2 - 2 \textbf{Re}(r)}{2(1 + |r|^2)}.
\end{equation}

\begin{figure}[!hbt]
\centering
\includegraphics[width=0.7\textwidth]{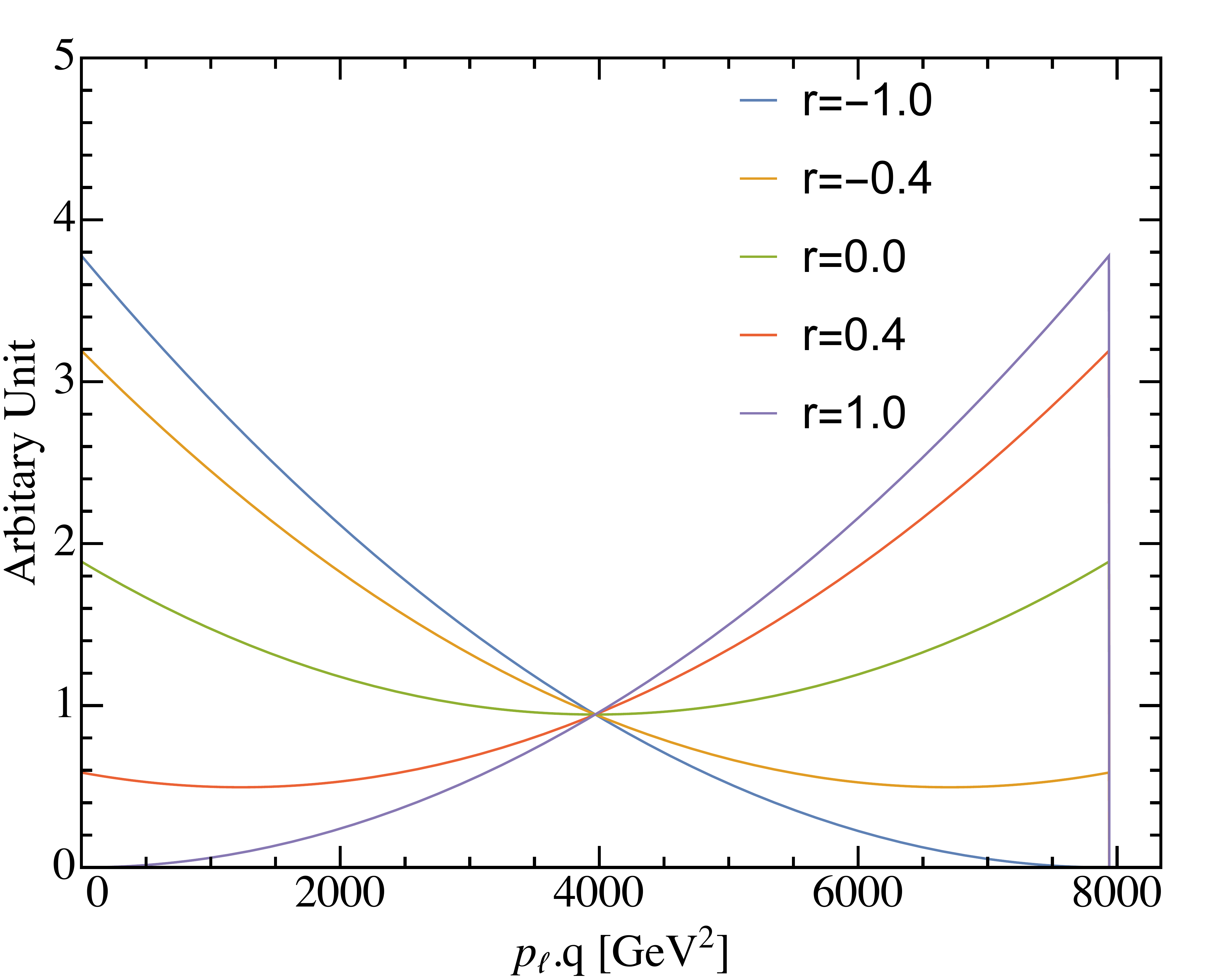}
\caption{The ideal $H^+ \to W^+ \gamma$ decay differential distribution of $p_\ell\cdot q$ for various values of $r \equiv \tilde S/S$.  For concreteness we set $m_{H^+} = 150$ GeV.}
\label{fig:WAME}
\end{figure}

We plot the ideal differential decay distribution in Fig.~\ref{fig:WAME} for various real values of $r$ between $-1$ and $+1$, as a function of the experimental observable $p_{\ell} \cdot q$.  Note that dividing Eq.~(\ref{eq:Msqr}) by an overall factor of $|r|^2$ yields the exact same distribution with $r \to 1/r$; therefore the differential distribution for real $r$ values outside the range $[-1,1]$ can be obtained trivially from Fig.~\ref{fig:WAME} by using this substitution.  For concreteness, we set $m_{H^+} = 150$~GeV; choosing different values of the charged Higgs mass only rescales the range of the $x$ axis in Fig.~\ref{fig:WAME}.

\subsection{Possible values of $r$}

We now consider the possible values that $r \equiv \tilde S/S$ can take.

The pseudoscalar form factor $\tilde S$ can be generated only by loops of fermions.  Therefore, for a purely fermiophobic charged Higgs, $r = 0$.  Phenomenologically, this is the most interesting situation because then the decays to light fermion pairs are absent and the branching ratio of $H^+ \to W^+ \gamma$ can be significant.  This is the case for $H_5^+$ of the GM model, which we will discuss further in the next section.

\begin{figure}
\resizebox{0.5\textwidth}{!}{\includegraphics{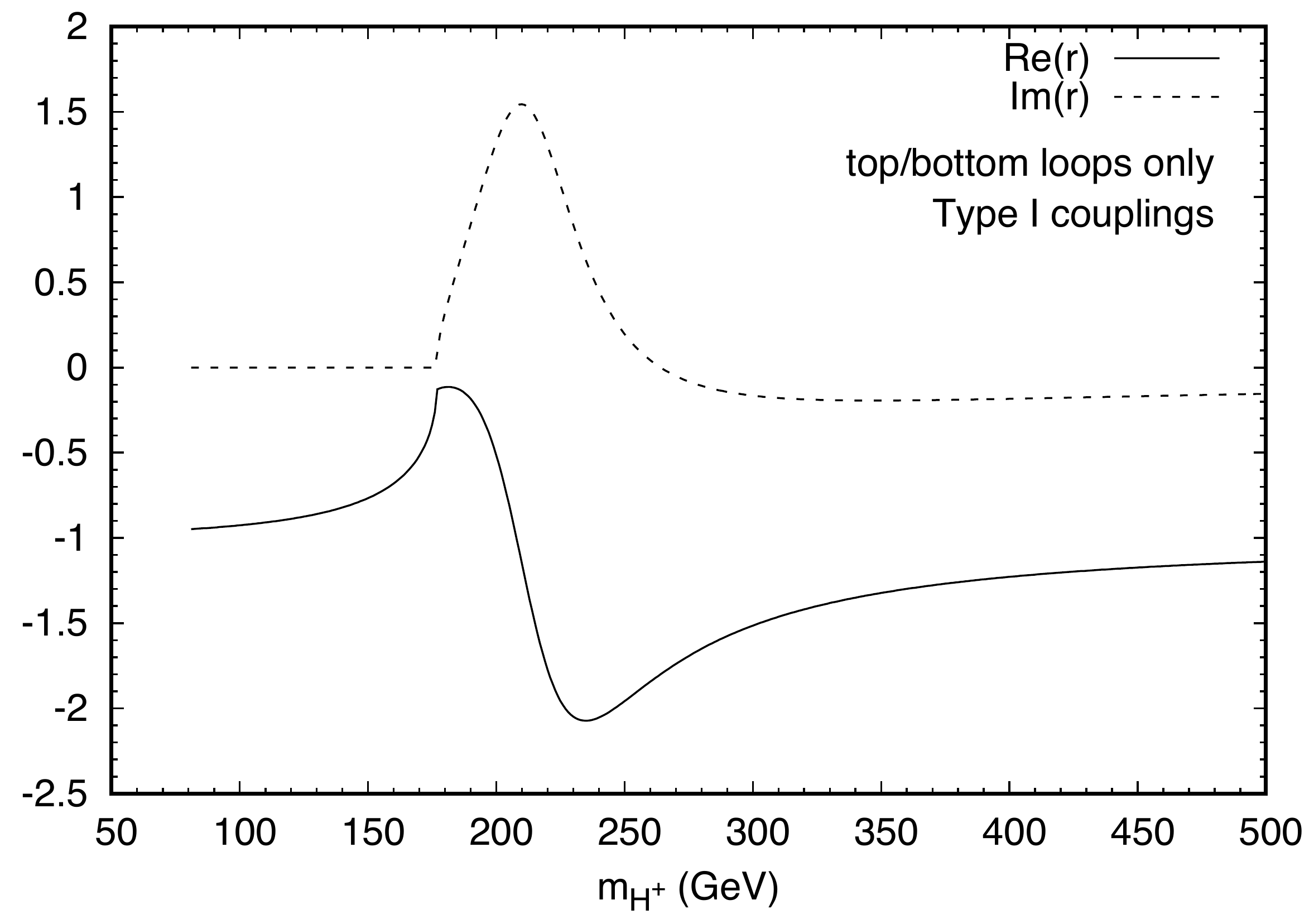}}
\resizebox{0.5\textwidth}{!}{\includegraphics{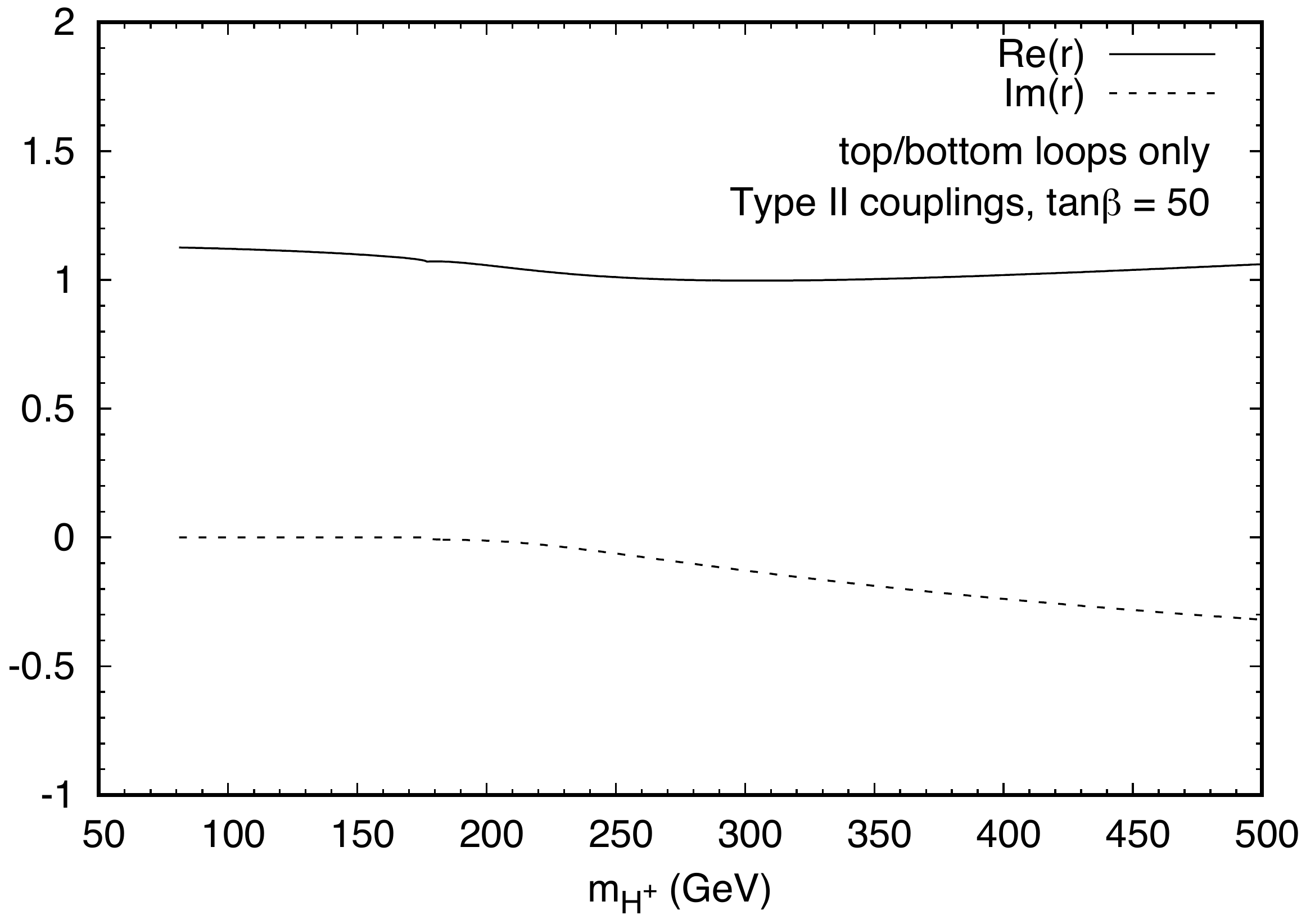}}
\caption{Real and imaginary parts of $r \equiv \tilde S/S$ including only the loops involving top and bottom quarks, for $H^+$ couplings as in the Type-I 2HDM or GM model (left) and as in the Type-II 2HDM with $\tan\beta = 50$ (right).}
\label{fig:rvalues}
\end{figure}

When $H^+$ is not fermiophobic, $\tilde S$ and $S$ both receive contributions from loops involving top and bottom quarks.  $S$ also generically receives contributions from loops involving scalars and/or gauge bosons.  Ignoring the bosonic loops, we can study the behaviour of $r$ due only to the top and bottom quark loops.  This is shown in Fig.~\ref{fig:rvalues}, where we implement only the top/bottom quark loop contributions to $S$ and $\tilde S$ using the calculation of Ref.~\cite{Degrande:2017naf} for the fermiophilic charged Higgs $H_3^+$ in the GM model.  The fermion couplings of $H_3^+$ follow the same pattern as in the Type-I 2HDM.  We also generalize to the Type-II 2HDM using the results of Ref.~\cite{Ilisie:2014hea} for the Aligned 2HDM, with the couplings as given in Table~\ref{tab:fcoups}~\cite{Pich:2009sp}.

\begin{table}
\begin{center}\begin{tabular}{lcc}
\hline\hline
Aligned 2HDM & $\zeta_u$ & $\zeta_d$ \\
Type-I 2HDM & $\cot\beta$ & $\cot\beta$ \\
GM model & $\tan\theta_H$ & $\tan\theta_H$ \\
Type-II 2HDM & $\cot\beta$ & $-\tan\beta$ \\
\hline\hline
\end{tabular}\end{center}
\caption{Charged Higgs couplings to fermions in the Aligned 2HDM, and corresponding values in the Type-I and -II 2HDMs and the GM model.  The couplings are defined in terms of the Feynman rule for the $H^+ \bar t b$ vertex, $-i \sqrt{2} [\zeta_d m_b P_R - \zeta_u m_t P_L]/v$, where $P_{R,L} = (1 \pm \gamma^5)/2$ and $v \simeq 246$~GeV is the SM Higgs vev.}
\label{tab:fcoups}
\end{table}

In the left panel of Fig.~\ref{fig:rvalues} we plot the real and imaginary parts of $r$ including the top/bottom quark loop only and taking the couplings of $H^+$ as in the Type-I 2HDM or the GM model.  Dependence on $\tan\beta$ or $\tan\theta_H$ cancels out in the ratio $r$, so $r$ depends only on the $H^+$ mass.  The threshold at which $H^+ \to t \bar b$ opens up is clearly visible.  Below this threshold, $r$ is real and lies between $-1$ and $0$.  Above this threshold, tree-level decays to $t \bar b$ compete with the loop-induced decay to $W^+ \gamma$, making the latter phenomenologically much less interesting.  

In the right panel of Fig.~\ref{fig:rvalues} we plot the real and imaginary parts of $r$ including the top/bottom quark loop only, this time taking the couplings of $H^+$ as in the Type-II 2HDM with $\tan\beta = 50$.  The threshold at which $H^+ \to t \bar b$ opens up is much less obvious, but still visible.  In this case, {\bf Re}($r$) is close to $+1$ over a wide range of $H^+$ masses.  $r$ now depends on the value of $\tan\beta$: Type-II couplings with $\tan\beta = 1$ lead to $r$ values nearly (but not exactly) identical to the left panel of Fig.~\ref{fig:rvalues}.

In a realistic model, $S$ also receives contributions from loops involving scalars and/or gauge bosons.  These can have either sign -- in particular, in the GM model with small $s_H$, the sign of the scalar loop contribution is controlled by the sign of the trilinear scalar coupling parameter $M_2$ [see Eq.~(\ref{equ:potential})].  Therefore, the scalar and/or gauge boson contributions to $S$ can interfere constructively or destructively with the fermion contribution, and can even change the sign of $S$.  This means that {\bf Re}($r$) can be larger or smaller in magnitude than shown in Fig.~\ref{fig:rvalues}, and can even change sign.

The general conclusion that we can draw from experimental detection of a nonzero value of $r$ from the shape of the $p_\ell \cdot q$ distribution is therefore rather limited: nonzero $r$ tells us only that the fermion loop contribution is non-negligible.  This implies that $H^+$ is not fermiophobic and can also be searched for via its fermionic decay products, and (for masses below the top quark mass) its production in top quark decays.

\section{A benchmark scenario}
\label{sec:benchmark}

For the remainder of this paper we adopt the GM model as a prototype in order to study in more detail the future LHC sensitivity to the $W\gamma$ decay channel of a fermiophobic charged Higgs.

\subsection{The Georgi-Machacek model}
\label{sec:GMdef}

The scalar sector of the GM model~\cite{Georgi:1985nv,Chanowitz:1985ug} consists of the usual SM complex scalar doublet $(\phi^+,\phi^0)^T$ with hypercharge\footnote{We use the convention $Q=T^3+Y/2$.} $Y=1$, together with a real triplet $(\xi^+,\xi^0,-\xi^{+*})^T$ with $Y=0$ and a complex triplet $(\chi^{++},\chi^+,\chi^0)^T$ with $Y=2$.
In order to avoid stringent constraints from the electroweak $\rho$ parameter, custodial symmetry is introduced by imposing a global SU(2)$_L\times$SU(2)$_R$ symmetry upon the scalar potential.  The isospin doublet is written as a bi-doublet under SU(2)$_L \times$SU(2)$_R$ and the two isospin triplets are combined into a bi-triplet in order to make the symmetry explicit,
\begin{eqnarray}
	\Phi = \left(\begin{matrix}
		\phi^{0*} & \phi^+ \\
		-\phi^{+*} & \phi^0
		\end{matrix} \right), \qquad 
	X=\left(\begin{matrix}
		\chi^{0*} & \xi^+ & \chi^{++} \\
		-\chi^{+*} & \xi^0 & \chi^+ \\
		\chi^{++*} & -\xi^{+*} & \chi^0
		\end{matrix} \right).
\end{eqnarray}
The vevs are given by
\begin{eqnarray}
	\langle\Phi\rangle=\frac{v_\phi}{\sqrt{2}}I_{2\times2},\qquad 
	\langle X\rangle=v_\chi I_{3\times3},
\end{eqnarray}
where $I$ is the unit matrix and the $W$ and $Z$ boson masses give the constraint,
\begin{eqnarray}
	v_\phi^2 + 8v_\chi^2\equiv v^2 = \frac{1}{\sqrt{2}G_F} \approx (246 \text{ GeV})^2.
\end{eqnarray}
The most general gauge-invariant scalar potential involving these fields that preserves custodial SU(2) is given, in the conventions of Ref.~\cite{Hartling:2014zca}, by:
\begin{eqnarray}
\label{equ:potential}
	V(\Phi,X)&=&\frac{\mu_2^2}{2}\Tr(\Phi^\dagger\Phi)+\frac{\mu_3^2}{2}\Tr(X^\dagger X)
	+\lambda_1[\Tr(\Phi^\dagger\Phi)]^2+\lambda_2\Tr(\Phi^\dagger\Phi)\Tr(X^\dagger X) 
	\nonumber \\
	&& + \lambda_3\Tr(X^\dagger X X^\dagger X)+\lambda_4[\Tr(X^\dagger X)]^2 
	- \lambda_5\Tr(\Phi^\dagger\tau^a\Phi\tau^b)\Tr(X^\dagger t^a Xt^b) \nonumber \\
	&& - M_1\Tr(\Phi^\dagger\tau^a\Phi\tau^b)(UXU^\dagger)_{ab} 
	- M_2\Tr(X^\dagger t^a Xt^b)(UXU^\dagger)_{ab},
\end{eqnarray}
where $\tau^a=\sigma^a/2$ and
\begin{eqnarray}
t^1 = \frac{1}{\sqrt{2}}\left(\begin{array}{ccc}
0 & 1 & 0 \\
1 & 0 & 1 \\
0 & 1 & 0 
\end{array}\right), \quad t^2=\frac{1}{\sqrt{2}}\left(\begin{matrix}
0 & -i & 0\\
i & 0 & -i\\
0 & i  & 0
\end{matrix}\right), \quad t^3 = \begin{pmatrix}
1 & 0 & 0\\
0 & 0 & 0\\
0 & 0 & -1
\end{pmatrix}.
\end{eqnarray}
The matrix $U$, which rotates $X$ into the Cartesian basis, is given by
\begin{eqnarray}
	U = \left(\begin{matrix}
	-\frac{1}{\sqrt{2}} & 0 & \frac{1}{\sqrt{2}}\\
	-\frac{i}{\sqrt{2}} & 0 & -\frac{i}{\sqrt{2}}\\
	0 & 1 & 0
	\end{matrix}\right).
\end{eqnarray}

The physical fields can be organized by their transformation properties under the custodial SU(2) symmetry into a fiveplet, a triplet and two singlets:
\begin{eqnarray}
\textbf{Fiveplet: }&& H^{++}_5 = \chi^{++},\quad 
	H^+_5 = \frac{\chi^+-\xi^+}{\sqrt{2}},\quad 
	H_5^0 = \sqrt{\frac{2}{3}}\xi^{0,r}-\sqrt{\frac{1}{3}}\chi^{0,r}, \nonumber \\
\textbf{Triplet: }&& H^+_3 = -s_H\phi^+ + c_H\frac{\chi^++\xi^+}{\sqrt{2}},\quad 
	H_3^0 = -s_H\phi^{0,i} + c_H \chi^{0,i}, \nonumber \\
\textbf{Singlets: }&& H_1^0 = \phi^{0,r},\quad 
	H_1^{0\prime} = \sqrt{\frac{1}{3}}\xi^{0,r} + \sqrt{\frac{2}{3}}\chi^{0,r},
\end{eqnarray}
where 
\begin{equation}
s_H \equiv \sin\theta_H = \frac{2\sqrt{2}v_\chi}{v}, \quad c_H \equiv \cos\theta_H = \frac{v_\phi}{v}.
\end{equation}

Within the fiveplet and triplet, the masses are degenerate at tree level, and are given in terms of the parameters of the scalar potential by
\begin{align}
m_5^2 &= \frac{M_1}{4v_\chi}v_\phi^2 + 12 M_2 v_\chi + \frac{3}{2}\lambda_5v_\phi^2 + 8\lambda_3 v_\chi^2, \nonumber \\
m_3^2 &= \frac{M_1}{4v_\chi}v^2 + \frac{\lambda_5}{2}v^2.
\end{align}

The two custodial singlets will mix by an angle $\alpha$ to give the two mass eigenstates $h$ and $H$,
\begin{align}
h = c_\alpha H_1^0 - s_\alpha H_1^{0\prime}, \nonumber \\
H = s_\alpha H_1^0 + c_\alpha H_1^{0\prime},
\end{align}
where $c_\alpha = \cos\alpha$ and $s_\alpha = \sin\alpha$. The mixing is controlled by the mass matrix,
\begin{equation}
\mathcal{M}^2 = \left(\begin{array}{cc}
\mathcal{M}_{11}^2 & \mathcal{M}_{12}^2 \\
\mathcal{M}_{12}^2 & \mathcal{M}_{22}^2 
\end{array}\right),
\end{equation}
where 
\begin{align}
\mathcal{M}_{11}^2 &= 8\lambda_1v_\phi^2, \nonumber \\
\mathcal{M}_{12}^2 &= \frac{\sqrt{3}}{2}v_{\phi}[-M_1 + 4(2\lambda_2 - \lambda_5)v_\chi], \nonumber \\
\mathcal{M}_{22}^2 &= \frac{M_1}{4v_\chi}v_\phi^2 - 6 M_2 v_\chi + 8(\lambda_3+3\lambda_4)v_\chi^2.
\end{align}

\subsection{Fermiophobic $H_5^{\pm}$ decays and parameter choices}

The custodial-fiveplet states $H_5$ have no doublet component, and hence are fermiophobic at tree level.  The fiveplet states do, however, couple at tree level to massive vector boson pairs with a coupling proportional to $s_H$.  They also take part in gauge couplings of the form $H_5 H_5 V$ and $H_5 H_3 V$, where $V = W$ or $Z$; in what follows we will assume that $m_5 < m_3$, in which case there are no decays of $H_5$ into other scalar states.  The remaining possible decay channels for the $H_5$ states are listed in Table~\ref{tab:H5Decay}, including the loop-induced decays involving one or more photons.

\begin{table}
\centering
\begin{tabular}{l|l|l}
\hline
\hline
Particle & Decay channels & Comment \\
\hline
$H_5^\pm$:\quad & $H_5^\pm \to W^\pm \gamma$ & Loop-induced\\
        & $H_5^\pm \to W^{\pm(*)} Z^{(*)}$ & Suppressed by $s_H^2$, off-shell\\
\hline
$H_5^0$:\quad   & $H_5^0 \to \gamma \gamma$ & Loop-induced\\
        & $H_5^0 \to Z \gamma$ & Loop-induced, phase space disfavored\\
        & $H_5^0 \to Z^{(*)}Z^{(*)}$ & Suppressed by $s_H^2$, off-shell\\
        & $H_5^0 \to W^{\pm(*)}W^{\mp(*)}$ & Suppressed by $s_H^2$, off-shell \\
\hline
$H_5^{\pm\pm}$:\quad & $H_5^{\pm\pm} \to W^{\pm(*)} W^{\pm(*)}$ & Suppressed by $s_H^2$, off-shell\\
\hline
\hline
\end{tabular}
\caption{Decay channels for members of the scalar fiveplet at low mass, including possible off-shell decays.  We consider the case in which the fiveplet is the lightest extra scalar; hence decays into other new scalars are kinematically forbidden.}
\label{tab:H5Decay}
\end{table}

\begin{figure}
\centering
\includegraphics[width=0.7\textwidth]{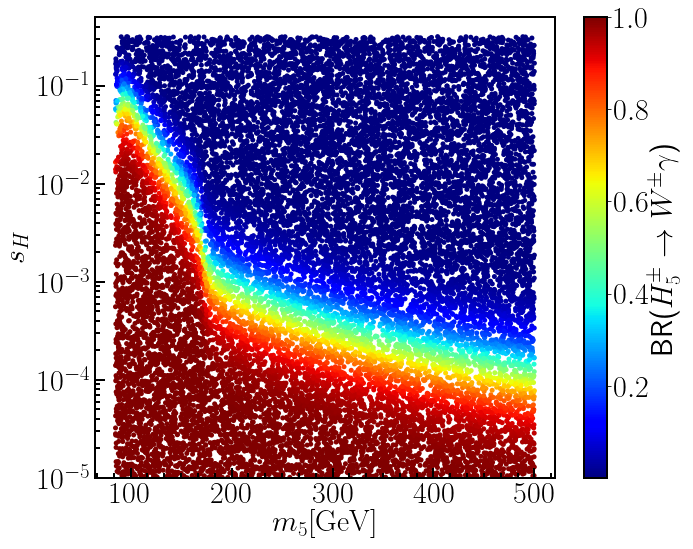}
\caption{Dependence of BR($H_5^\pm\to W^\pm\gamma$) on $m_{5}$ and $s_H$, for $M_2 = 40$~GeV.  Larger values of $M_2$ would move the transition to large BR($H_5^\pm\to W^\pm\gamma$) upwards to higher $s_H$ values, and smaller values of $M_2$ would move this transition to lower $s_H$ values.}
\label{fig:BRWA_thvsm5}
\end{figure}

The decay width for $H_5^{\pm} \to W^{\pm} \gamma$ is naturally small because this process is loop suppressed.  This decay therefore can become important only when the competing tree-level $H_5^{\pm} \to W^{\pm}Z$ decay is sufficiently suppressed.  This can happen in two ways: (i) when $s_H$ is small, suppressing the $H_5^{\pm} W^{\mp} Z$ coupling; and/or (ii) when $m_5$ is below the $WZ$ threshold, where the $H_5^{\pm} \to W^{\pm} Z$ decay is off-shell and hence kinematically suppressed.  These two parameter regions are illustrated in Fig.~\ref{fig:BRWA_thvsm5}, where we show the dependence of BR($H_5^{\pm} \to W^{\pm} \gamma$) on $m_5$ and $s_H$, taking $M_2 = 40$~GeV and fixing the other parameters according to (see Appendix~\ref{app:LSH})\footnote{We will adopt the choice of parameters in Eq.~(\ref{eq:params}) for the remainder of this paper, keeping $m_5$, $s_H$, and $M_2$ as free parameters whose values we will specify.}
\begin{align}
	m_3^2 &= m_5^2 + \delta m^2, \nonumber \\
	m_H^2 &= m_5^2 + \frac{3}{2}\delta m^2 + \kappa_Hv^2s_H^2, \nonumber \\
	M_1 &= \frac{\sqrt{2}}{v}\left(m_5^2+\frac{3}{2}\delta m^2\right)s_H + 3M_2s_H^2 + \kappa_{\lambda_3}vs_H^3, \nonumber \\
	s_\alpha &= \kappa_\alpha s_H,\nonumber \\
	\delta m^2 &= (300 \text{ GeV})^2, \nonumber \\
	\kappa_\alpha &= -0.15 - \frac{m_5}{1000 \text{ GeV}},\nonumber \\
	\kappa_H &= - \frac{m_5}{100 \text{ GeV}}, \nonumber \\
	\kappa_{\lambda_3} &= -\frac{\kappa_H^2}{10}.
\label{eq:params}
\end{align}
This choice of parameters ensures that the full range of $m_5$ and $s_H$ shown in Fig.~\ref{fig:BRWA_thvsm5} satisfies the theoretical constraints from perturbative unitarity of two-to-two scalar scattering amplitudes, boundedness from below of the potential, and the absence of deeper alternative minima~\cite{Hartling:2014zca}.
The kinematic threshold below which the competing $H_5^{\pm} \to W^{\pm}Z$ channel goes off shell is clearly visible.  Guided by this, we will concentrate on the region with $m_5 < 200$~GeV and $s_H$ fairly small.  We have chosen $m_3$ and $m_H$ to be large so that we can (conservatively) ignore their contributions to $H_5^{\pm}$ production, which we discuss in the next subsection.

The amplitude for the loop-induced decay $H_5^{\pm} \to W^{\pm} \gamma$ receives contributions from loop diagrams involving charged scalars $H_5^{\pm, \pm\pm}$ and $H_3^{\pm}$, $W$ and $Z$ bosons, and mixed diagrams involving both scalars and gauge bosons~\cite{Degrande:2017naf}.  The amplitudes for the gauge and mixed loop diagrams are all proportional to $s_H$, and hence are suppressed when $s_H$ is small.  This leaves the diagrams involving scalars in the loop, which are not suppressed at small $s_H$.  Instead, at small $s_H$, these diagrams are all proportional to the trilinear scalar coupling parameter $M_2$, and depend also on the masses $m_5$ and $m_3$ of the scalars in the loop (details are given in Appendix~\ref{app:FormFactor}).  With our choice $m_3 \gg m_5$, the loops involving $H_3^{\pm}$ become small, and the partial width for $H_5^{\pm} \to W^{\pm} \gamma$ essentially becomes a function of only $m_5$ and $M_2$ at small $s_H$.  The partial width for the competing tree-level decay $H_5^{\pm} \to W^{\pm}Z$ is proportional to $s_H^2$.  Thus, for a given mass $m_5$ and $s_H$ not too large, the branching fractions of $H_5^{\pm}$ are determined entirely by $s_H$ and $M_2$.

\subsection{$H_5^{\pm}$ production processes}
\label{sec:production}

Because the $H_5$ states are fermiophobic, we focus on gauge-boson-initiated production processes.
The relevant interactions of $H_5$ with one or two gauge bosons have the following coupling strengths:
\begin{align}
&g_{H_5^+H_5^-\gamma} = e, & g_{H_5^+H_5^-Z} = \frac{e}{2s_Wc_W}(1-2s^2_W), \nonumber \\
&g_{H_5^+H_5^{--}W^+} = \frac{e}{\sqrt{2} s_W}, & g_{H_5^+H_3^-Z} = -\frac{e}{2s_Wc_W}c_H, \nonumber \\
&g_{H_5^+H_5^0W^-} = \frac{\sqrt{3}e}{2s_W}, & g_{H_5^+H_3^0W^-} = -\frac{ie}{2s_W}c_H, \nonumber \\
&g_{H_5^+W^-Z} = -\frac{e^2v}{2s^2_Wc_W}s_H, & g_{H_5^0W^+W^-} = \frac{e^2v}{2 \sqrt{3} s_W^2}s_H, \nonumber \\
&g_{H_5^0ZZ} = -\frac{e^2v}{\sqrt{3} s_W^2c_W^2}s_H, \ \ & g_{H_5^{++}W^-W^-} = \frac{e^2v}{\sqrt{2} s_W^2}s_H.
\label{equ:H5Coupling}
\end{align}
Note that all the couplings of $H_5$ to two gauge bosons are proportional to $s_H$, while the couplings of two scalars ($H_5 H_5$ or $H_5 H_3$) to one gauge boson are either a gauge coupling or a gauge coupling times $c_H$.  Therefore for $s_H \ll 1$, the cross sections for single $H_5$ production (via vector boson fusion or associated production with a vector boson) will be suppressed by $s_H^2$, while Drell-Yan processes that produce a pair of $H_5$ states (or $H_5 H_3$) will be unsuppressed, with cross sections controlled only by the relevant gauge coupling and the masses of the final-state scalars.  

\begin{figure}
\centering
\includegraphics[width=0.95\textwidth]{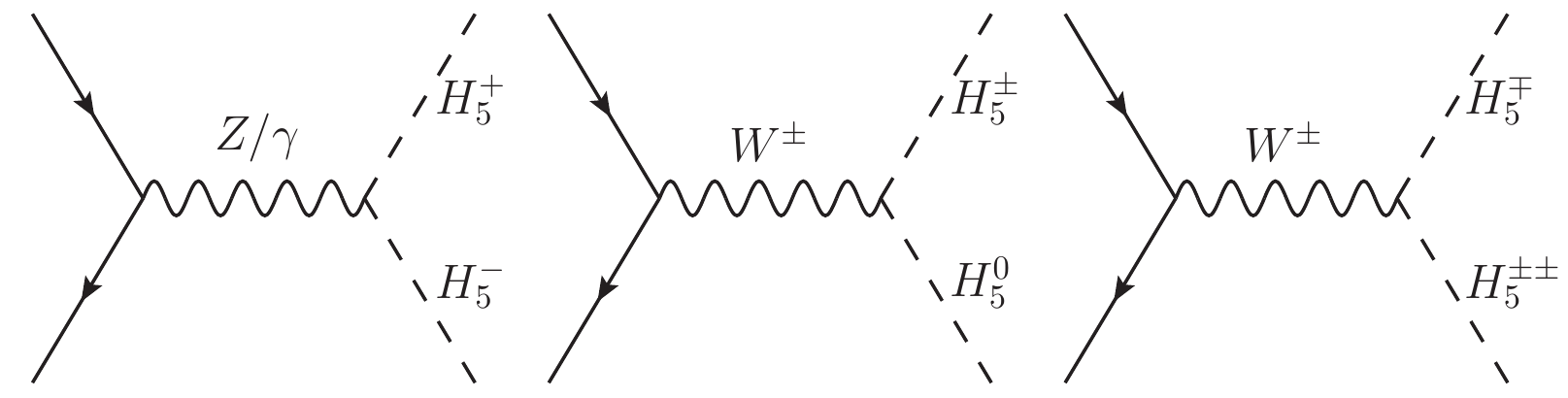}
\caption{Feynman diagrams for the dominant Drell-Yan production processes involving $H_5^\pm$ when $s_H \ll 1$ and $m_3 \gg m_5$.}
\label{fig:DYproduction}
\end{figure}

\begin{figure}
\centering
\includegraphics[width=0.70\textwidth]{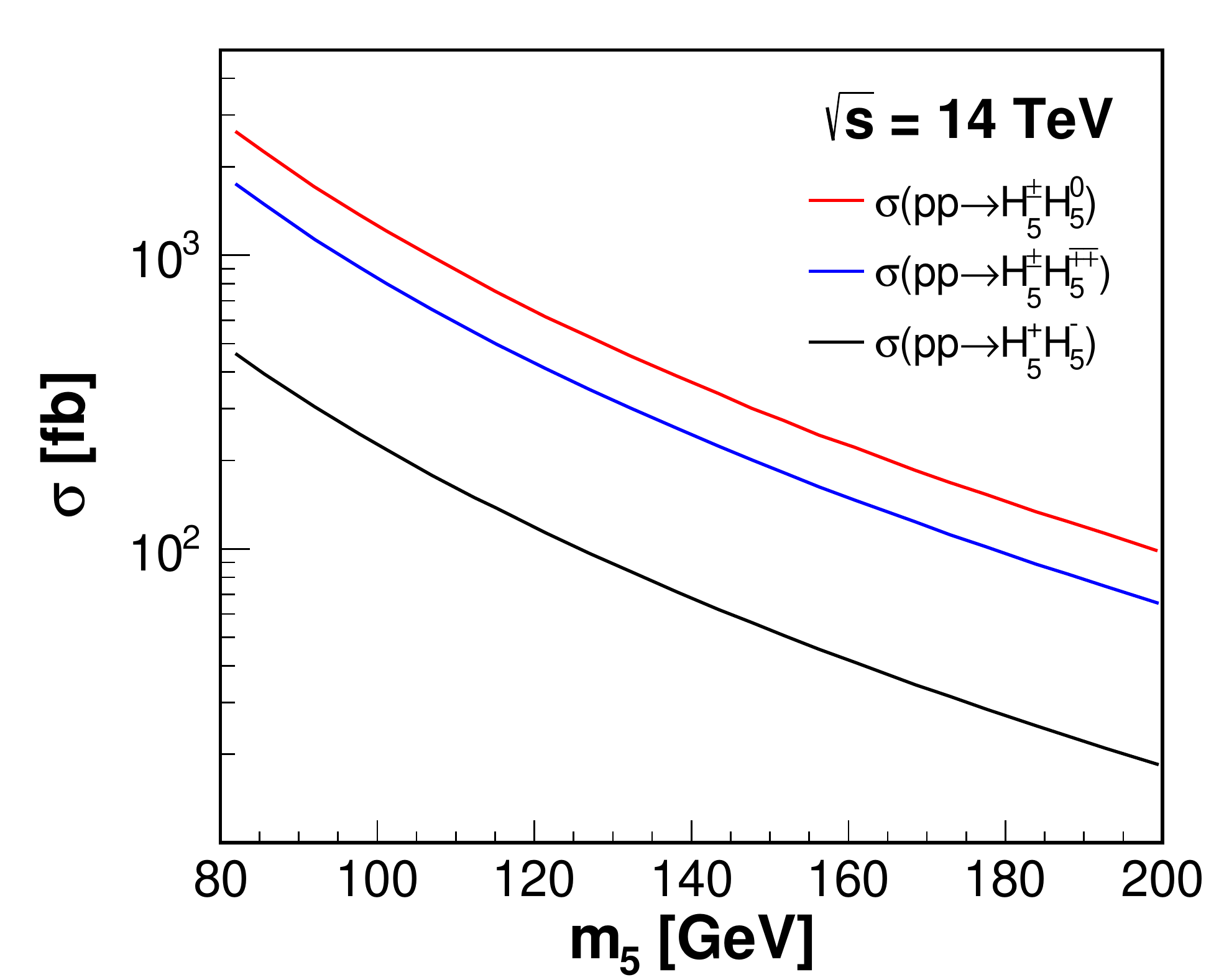}
\caption{Leading-order cross sections for the Drell-Yan production processes involving $H_5^\pm$ and another $H_5$ scalar, for $pp$ collisions at a centre-of-mass energy of 14~TeV.}
\label{fig:H5CS}
\end{figure}

Taking $m_3 \gg m_5$, we can ignore the contribution from associated $H_5 H_3$ production.\footnote{We also ignore the possible contribution from $q \bar q, gg \to H \to H_5^+ H_5^-$.  The $q \bar q H$ coupling (which also controls $gg \to H$) is suppressed in the small-$s_H$ limit.}  The most important production channels for $H_5^{\pm}$ are then $pp \to H_5^{\pm} H_5^0$, $H_5^{\pm} H_5^{\mp\mp}$, and $H_5^+ H_5^-$.  The Feynman diagrams are shown in Fig.~\ref{fig:DYproduction}.  These cross sections depend only on $m_5$, as illustrated in Fig.~\ref{fig:H5CS} for $\sqrt{s} = 14$~TeV and $m_5$ between 80 and 200~GeV.  These are calculated at leading order in QCD with MadGraph5-2.4.3~\cite{Alwall:2014hca}, using the NNPDF23 parton distribution set~\cite{Ball:2013hta} and the model implementation described in the next section.  $pp \to H_5^{\pm} H_5^0$ has the largest cross section, reaching above a picobarn for $m_5 = 100$~GeV.  The cross section for $pp \to H_5^{\pm} H_5^{\mp\mp}$ is smaller by a factor of $2/3$, due entirely to the different couplings in Eq.~(\ref{equ:H5Coupling}).  The smallest is $pp \to H_5^+ H_5^-$, reaching a little over 200~fb for $m_5 = 100$~GeV.  While these Drell-Yan cross sections drop rapidly with increasing $m_5$, they offer plenty of events at low mass if the signal is sufficiently clean.

\section{Search prospects at the LHC}
\label{sec:search}

We now study the search prospects for the charged Higgs in the $W \gamma$ channel.  We focus on the mass range $m_5 \in (80, 200)$~GeV and project the exclusion reach for 300~fb$^{-1}$ at the 14~TeV LHC.

\subsection{Model implementation}

The whole GM model at leading and next-to-leading orders in QCD has previously been implemented in FeynRules~\cite{Alloul:2013bka} and a UFO~\cite{Degrande:2011ua} model file produced for simulation purposes.  We extend the leading order FeynRules implementation to include effective vertices of the form given in Eq.~(\ref{eq:vertex}) for all loop-induced decays of the scalars into gauge boson pairs that are not present at tree level~\cite{GMUFO}.  The one-loop calculations of these effective vertices were already implemented in GMCALC 1.3.0 for the purpose of calculating decay branching ratios; we adapt GMCALC to write the effective coupling form factors in a param\_card.dat file for use by MadGraph5~\cite{Alwall:2014hca}.  (This adaptation is included in the public release of GMCALC 1.4.0.)  This implementation allows us to accurately simulate the kinematics of the loop-induced scalar decays.

\subsection{Simulation and selection cuts}

In order to determine the sensitivity of a charged scalar search in the $W\gamma$ channel, we perform a cut-based Monte Carlo analysis of the inclusive $W\gamma$ signal.  In particular, we require at least one lepton ($e^{\pm}$ or $\mu^{\pm}$) and at least one photon in the final state.  Signal and background events are generated at leading order in QCD using MadGraph5~\cite{Alwall:2014hca}, showered and hadronized using Pythia~\cite{Sjostrand:2014zea,Sjostrand:2006za}, and then passed to Delphes~\cite{deFavereau:2013fsa} for the detector simulation.  

The signal processes, as discussed in Sec.~\ref{sec:production}, are
\begin{align}
p p\ &\to H_5^\pm H_5^0 \to W^\pm \gamma + X \to \ell \ \nu_\ell \ \gamma + X, \nonumber \\
p p\ &\to H_5^\pm H_5^{\mp\mp}\to W^\pm \gamma + X \to  \ell \ \nu_\ell \ \gamma + X, \nonumber \\
p p\ &\to H_5^+ H_5^- \to W^\pm \gamma + X \to \ell \ \nu_\ell \ \gamma + X.
\end{align}
We generate the inclusive signal requiring at least one lepton and at least one photon (with kinematic requirements given below).  While we will vary BR($H_5^{\pm} \to W^{\pm} \gamma$) in order to extract limits on this branching ratio, we have to make some assumptions about the decay branching ratios of the other $H_5$ states produced in association.  In our simulation we assume that ${\rm BR}(H_5^{\pm\pm}\to W^\pm W^\pm) = 1$ and ${\rm BR}(H_5^0\to\gamma\gamma) = 1$. The first of these is a safe assumption because this is the only possible two-body decay of $H_5^{\pm\pm}$ when $m_3 > m_5$.  The second is a conservative assumption because the additional photons from $H_5^0$ introduce combinatoric background and reduce the signal efficiency. Finally, for the $H_5^+ H_5^-$ channel, we allow the second $H_5^{\pm}$ to decay into either $W^{\pm}\gamma$ or $W^{\pm}Z$, taking ${\rm BR}(H_5^\pm\to W^\pm Z) = 1-{\rm BR}(H_5^\pm\to W^\pm\gamma)$.  Again, this is a safe assumption so long as $m_3 > m_5$.

We simulate the following SM processes as backgrounds:
\begin{align}
p p\ &\to W^\pm \gamma \to \ell\ \nu_\ell\ \gamma, \nonumber \\
p p\ &\to W^\pm \gamma \gamma \to \ell\ \nu_\ell\ \gamma\ \gamma, \nonumber \\
p p\ &\to W^+ W^- \gamma \to \ell\ \nu_\ell\ \gamma + X, \nonumber \\
p p\ &\to W^+ W^- \gamma \gamma\to \ell\ \nu_\ell\ \gamma\ \gamma + X, \nonumber \\
p p\ &\to t \bar{t} \gamma \to \ell\ \nu_\ell\ \gamma + X, \nonumber \\ 
p p\ &\to W^\pm Z \gamma \to \ell\ \nu_\ell\ \gamma +X.
\end{align}
$W^\pm\gamma$ has the largest cross section before cuts, but it can be easily suppressed by the cuts described below.  The dominant background after cuts is $t \bar t \gamma$, followed by $W^+W^-\gamma$ and $W^+W^-\gamma\gamma$.  When calculating the signal significance, we include an overall 10\% systematic error on the background cross section.  We do not include fake backgrounds, which we feel are best estimated by experimentalists, for example, through data-driven methods.  These could reduce the sensitivity to our signal.

We begin by requiring at least one lepton with transverse momentum $p_T > 25$ GeV and pseudorapidity $|\eta|<2.5$ and at least one photon with $p_T > 25$ GeV and $|\eta|<2.5$.
To reduce combinatoric backgrounds from mis-pairings of the lepton and photon in signal events, we take the following strategy.
When more than one lepton passes the $p_T$ and $\eta$ requirements, we choose the highest-$p_T$ lepton as most likely to have come from the decay of $H_5^{\pm}$.  This is mostly an issue for the $pp \to H_5^{\pm} H_5^{\mp\mp}$ signal process; because the $H_5^{\mp\mp}$ must decay to two $W$ bosons, they are more likely to be off-shell than the $W$ from $H_5^{\pm} \to W^{\pm} \gamma$, and hence their decay leptons are generally softer.  When more than one photon passes the $p_T$ and $\eta$ requirements, we choose the photon with the smallest separation $\Delta R \equiv \sqrt{(\Delta \eta)^2 + (\Delta \phi)^2}$ (where $\Delta \phi$ is the azimuthal separation in radians) from our chosen lepton.  This is mostly an issue for $pp \to H_5^{\pm} H_5^0$ with $H_5^0 \to \gamma\gamma$, as well as for $pp \to H_5^+ H_5^-$ when both charged Higgs bosons decay to $W\gamma$.  Because the Drell-Yan scalar pair production process is $p$-wave, the scalars tend to be somewhat boosted, making the selection based on $\Delta R$ sufficiently effective.\footnote{Choosing the photon with highest $p_T$ is \emph{not} a good strategy, because the photons from $H_5^0 \to \gamma\gamma$ tend to have higher $p_T$ than the photon from $H_5^{\pm} \to W^{\pm}\gamma$.}

\begin{figure}
\centering
\includegraphics[width=0.48\textwidth]{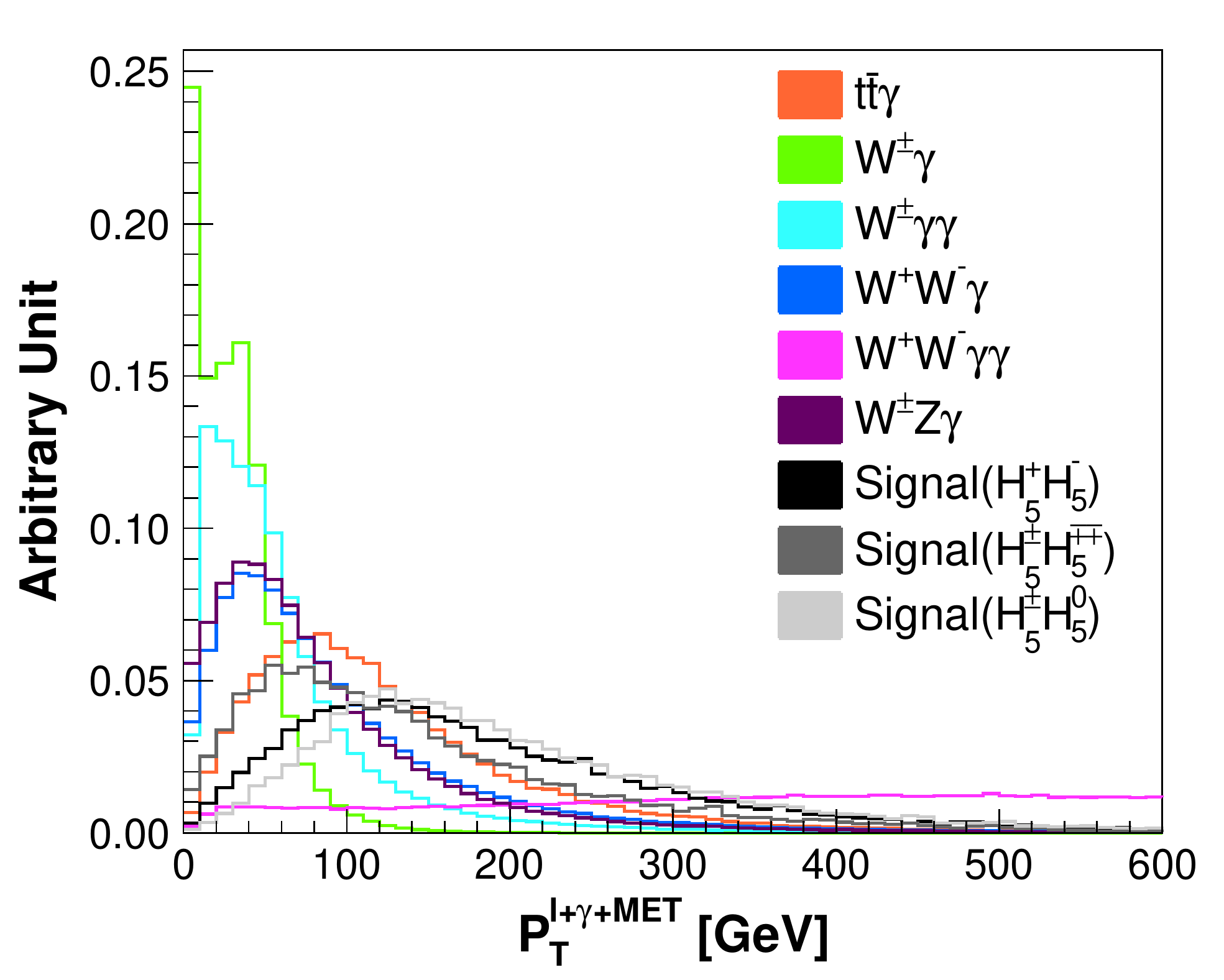}
\includegraphics[width=0.48\textwidth]{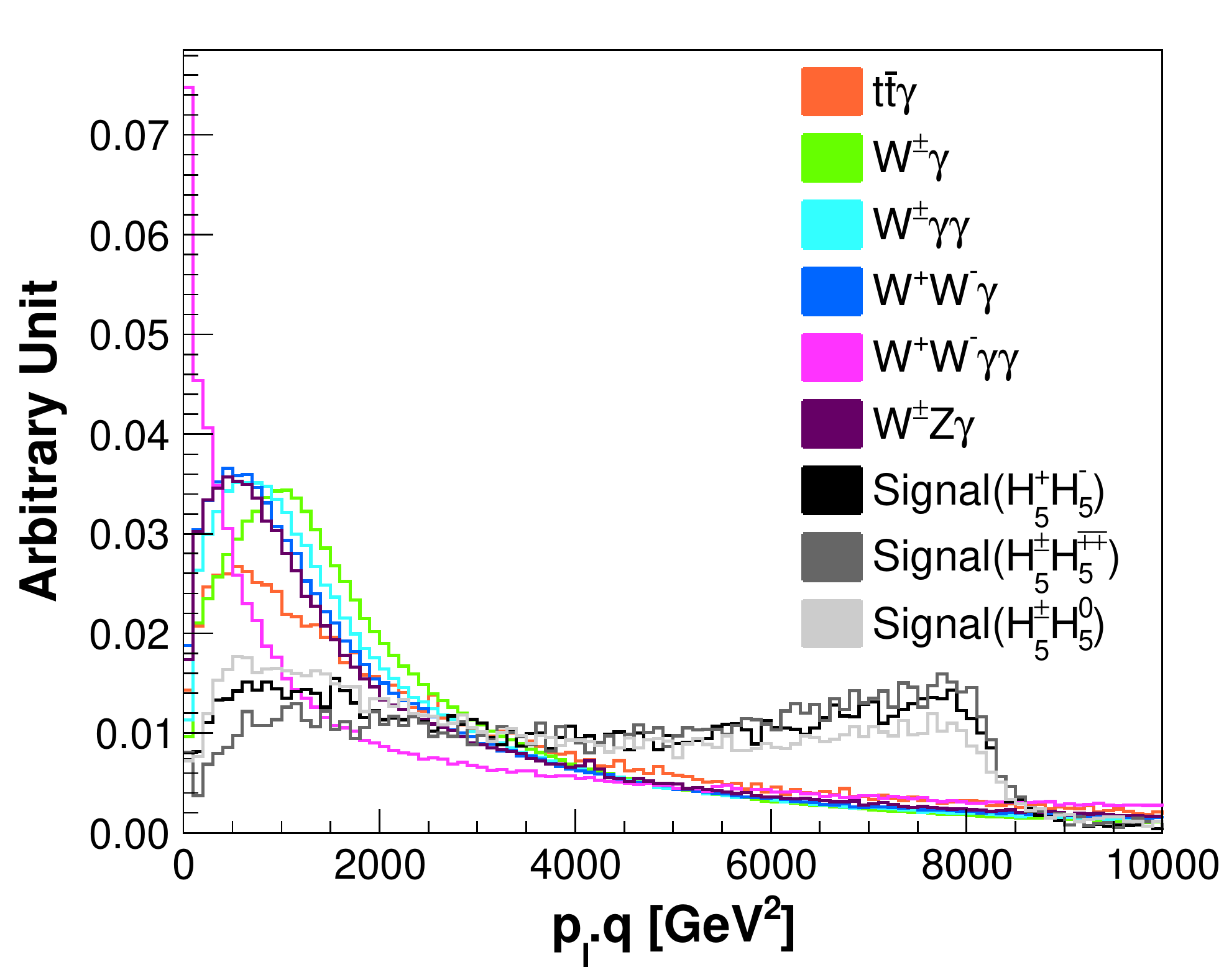}
\caption{Normalized distributions of $p_T^{\ell+\gamma+\slashed{E}_T}$ (left) and $p_\ell\cdot q$ (right) for $m_{5} = 150$ GeV for the signal and background processes.  The characteristic peak in $p_\ell \cdot q$ at the kinematic endpoint at $(m_5^2 - m_W^2)/2$ is visible in the right plot.  The deviation of the $p_{\ell} \cdot q$ distribution from the ideal parabolic shape at low $p_{\ell} \cdot q$ is mainly due to mis-pairing of the lepton and photon.}
\label{fig:distributions}
\end{figure}

We then apply additional cuts on each of the following variables:
\begin{itemize}
\item $N_j$, the number of reconstructed jets with $p_T > 20$~GeV, and $N_b$, the number of the jets that are tagged as $b$ jets by Delphes; in all cases we require $N_j \leq 2$ and $N_b=0$.  This helps to reduce the $t\bar{t}\gamma$ background;
\item $\slashed{E}_T$, the missing transverse energy;
\item $H_T$, the scalar sum of the $p_T$ of all visible objects;
\item $p_T^{\ell+\gamma+\slashed{E}_T}$, the vector sum of the $p_T$ of our chosen lepton and photon together with the missing transverse momentum.  In events with only one neutrino, this is equal to the transverse momentum of $H_5^{\pm}$;
\item $p_\ell \cdot q$, the dot product of the four-momenta of our chosen lepton and photon, which was identified as a useful variable in Sec.~\ref{sec:differential}.
\end{itemize}
The distributions of the last two variables for each signal and background process are shown in Fig.~\ref{fig:distributions} for $m_5 = 150$~GeV.

The cuts are optimized for the best signal significance for each value of $m_5$.\footnote{Note that when $m_5$ is close to $m_W$, the photon coming from $H_5^{\pm} \to W^{\pm} \gamma$ becomes soft and the parton-level upper limit of $p_\ell\cdot q$ becomes close to zero, making reconstruction of the correct lepton and photon difficult and leading to numerical instabilities in the automatic optimization of the cuts.  To avoid this, for $m_5<100$~GeV we fix the cuts at the values obtained for $m_5 =100$~GeV. } For example, for $m_5 = 150$~GeV, we take
\begin{itemize}
\item $72\text{ GeV}\leq\slashed{E}_T \leq 220$ GeV,
\item $260 \text{ GeV}<H_T< 620 \text{ GeV}$,
\item $100\text{ GeV} <p_T^{\ell+\gamma+\slashed{E}_T}< 420 \text{ GeV}$, 
\item $3300\text{ GeV}^2 < p_\ell\cdot q < 8200\text{ GeV}^2$.
\end{itemize}
The expected cross section of each signal and background process before and after applying these cuts is listed in Table~\ref{tab:csafter} for $m_5 = 150$~GeV assuming ${\rm BR}(H_5^\pm\to W^\pm\gamma)=1$ for the signal processes.  

\begin{table}
\centering
\resizebox{\textwidth}{!}{
\begin{tabular}{|c|ccc|cccccc|}
\hline
Process & $H_5^\pm H_5^0$ & $H_5^\pm H_5^{\mp\mp}$ & $H_5^+H_5^-$ & $t\bar{t}\gamma$ & $W^\pm\gamma$ & $W^\pm\gamma\gamma$ & $W^+W^-\gamma$ & $W^+W^-\gamma\gamma$ & $W^\pm Z\gamma$ \\
\hline
$\sigma \times$BR [fb] (before cuts) & 57.29 & 38.19 & 19.07 & 856 & 23000 & 30 & 120 & 65 & 25 \\
$\epsilon \times \sigma \times$BR [fb] (after cuts) & 4.21 & 1.01 & 0.95 & 0.49 & 0.09 & 0.05 & 0.38 & 0.28 & 0.05 \\
\hline
\end{tabular}
}
\caption{The cross section times branching ratio of each process before and after applying the cuts for $m_5 = 150$~GeV, defined as for the fiducial cross section in Eq.~(\ref{eq:fiducial}). For the signal processes, we assume ${\rm BR}(H_5^\pm\to W^\pm\gamma)=100\%$ and use ${\rm BR}(W^\pm\to\ell^\pm\nu)\approx21.34\%$.  
}
\label{tab:csafter}
\end{table}

\begin{figure}
\centering
\includegraphics[width=0.45\textwidth]{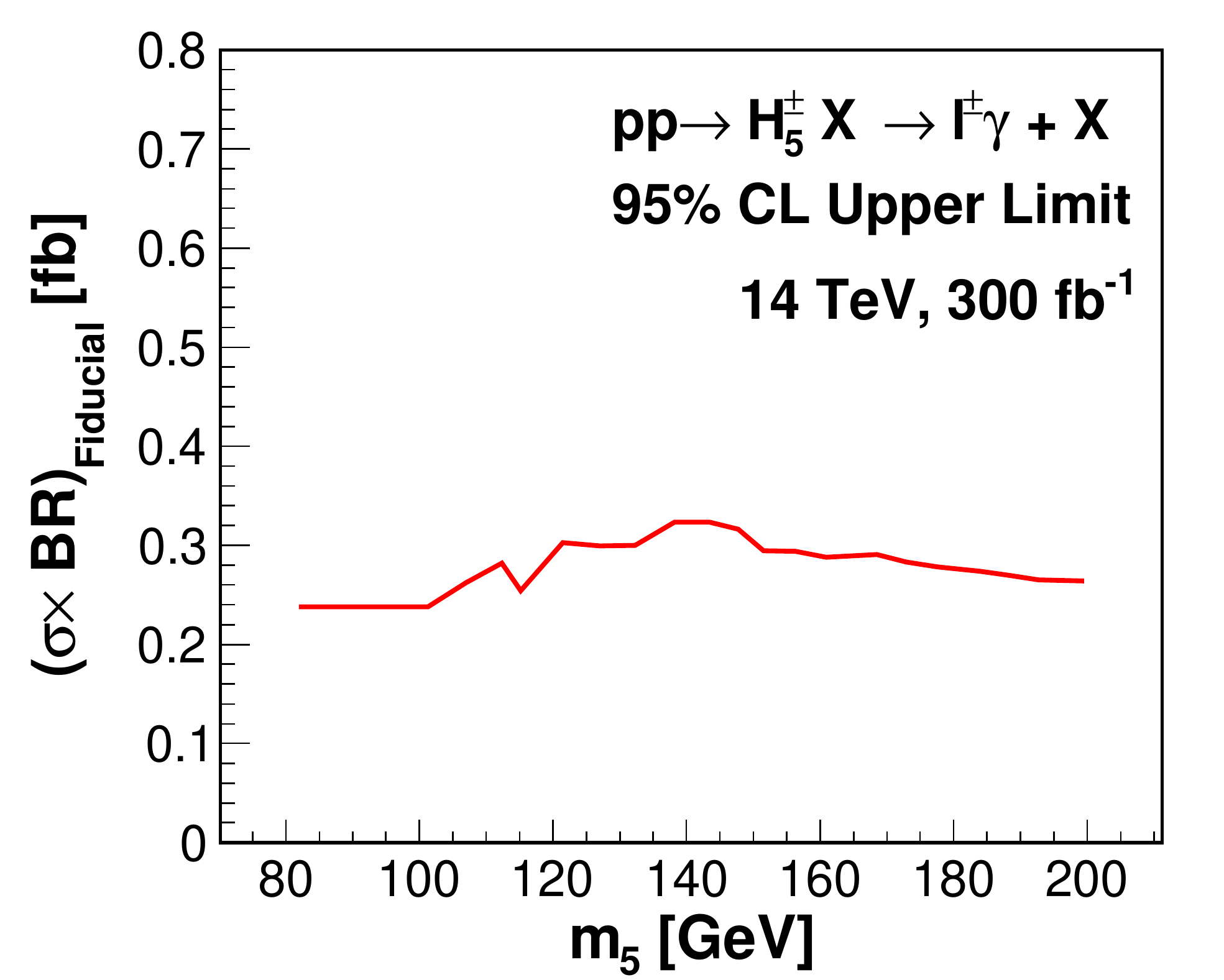}
\includegraphics[width=0.45\textwidth]{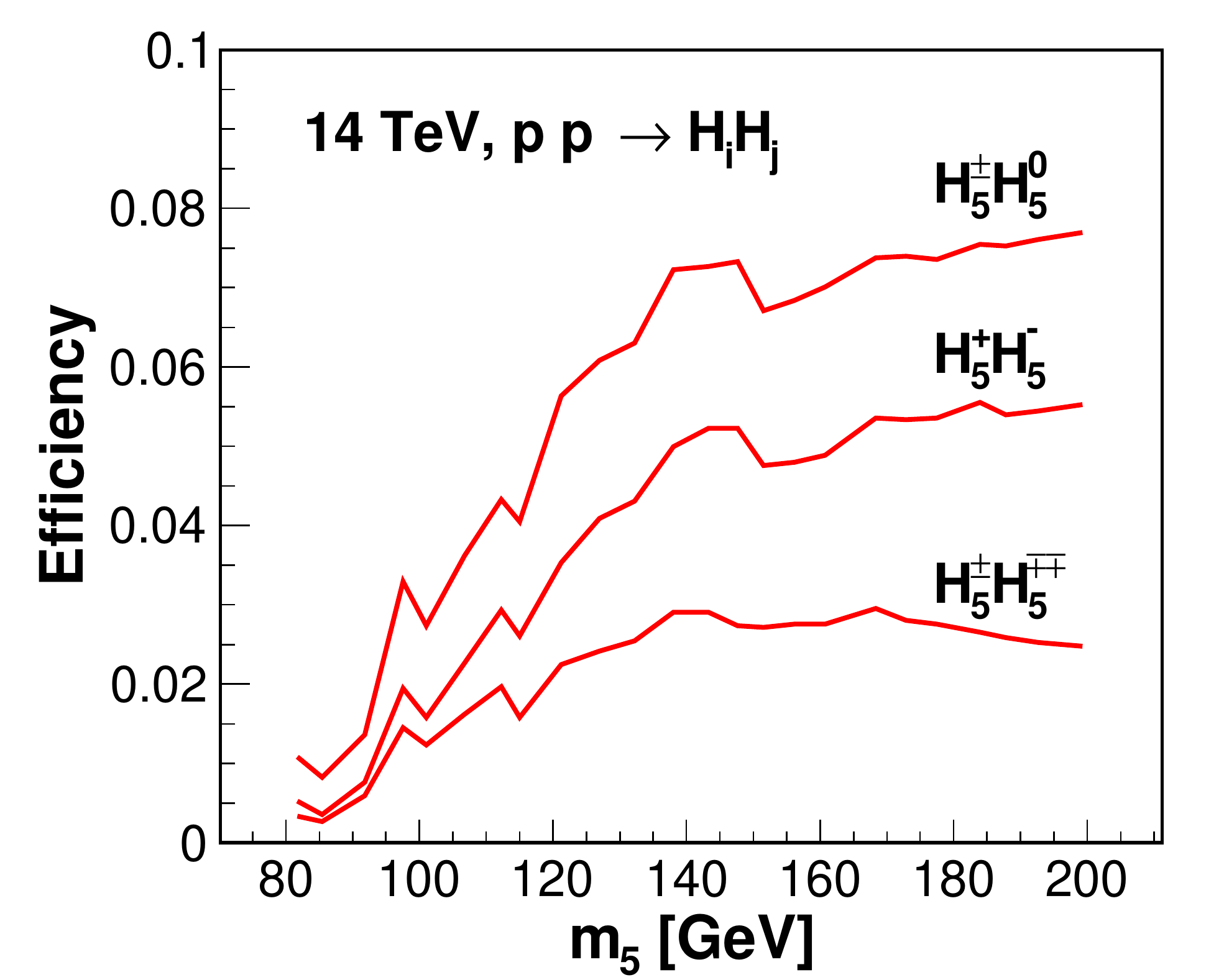}\\
\caption{The projected 95\% confidence level (CL) upper limit on $(\sigma\times{\rm BR})_{\text{Fiducial}}$ with 300~fb$^{-1}$ of data at the 14~TeV LHC (left panel) and the efficiency $\epsilon$ of the cuts that define the fiducial volume for each signal process (right panel).}
\label{fig:CSFiducialUpperlimit}
\end{figure}

Because each production process has a different efficiency to pass the cuts and because the contribution to the signal rate of the $H_5^+ H_5^-$ process depends nonlinearly on BR($H_5^{\pm} \to W^{\pm} \gamma$), we first present the expected upper limit on the fiducial cross section as a function of $m_5$ in the left panel of Fig.~\ref{fig:CSFiducialUpperlimit}.  The fiducial cross section is defined as
\begin{align}
&(\sigma\times{\rm BR})_{\text{Fiducial}} \equiv \epsilon_{H_5^{\pm} H_5^0} \sigma(pp \to H_5^{\pm} H_5^0){\rm BR}(H_5^{\pm} \to \ell^\pm\nu\gamma)  \nonumber \\
&\quad+ \epsilon_{H_5^{\pm} H_5^{\mp\mp}}\sigma(pp \to H_5^{\pm} H_5^{\mp\mp}){\rm BR}(H_5^{\pm} \to \ell^\pm\nu\gamma) \nonumber \\
&\quad+\epsilon_{H_5^{+} H_5^{-}}\sigma(pp \to H_5^+ H_5^-)\left[2 {\rm BR}(H_5^{\pm} \to \ell^\pm\nu\gamma) - {\rm BR}(H_5^{\pm} \to \ell^\pm\nu\gamma)^2\right].
\label{eq:fiducial}
\end{align}
Here ${\rm BR}(H_5^{\pm} \to \ell^{\pm} \nu \gamma) = {\rm BR}(H_5^{\pm} \to W^{\pm}\gamma) \times {\rm BR}(W^{\pm} \to \ell^{\pm} \nu)$ and $\epsilon_{H_iH_j}$ stands for the efficiency of the cuts for the process $pp \to H_i H_j$.  This efficiency is shown for each signal process in the right panel of Fig.~\ref{fig:CSFiducialUpperlimit}. 
As the mass of the scalar approaches the threshold of the $W\gamma$ channel, the efficiency drops to near zero. This is due to the photon becoming too soft to pass the initial selection as well as the variable $p_\ell\cdot q$ losing its discriminative ability when $m_{5}$ is close to $m_W$.  The upturn in the efficiency for $m_5 \sim m_W$ in the right panel of Fig.~\ref{fig:CSFiducialUpperlimit} is due to a (counterintuitive) rise in the number of photons passing the minimum $p_T$ threshold in our simulation as the $W$ is pushed off shell.  Because the form factor for the $H^{\pm} W^{\mp} \gamma$ vertex that we use in our calculation is computed assuming on-shell external particles, we will consider our results reliable only for $m_5 \gtrsim 100$~GeV.  As we will see in Sec.~\ref{sec:gaga}, lower $m_5$ values are mostly well covered by searches for $H_5^0 \to \gamma\gamma$.

\begin{figure}
\centering
\includegraphics[width=0.45\textwidth]{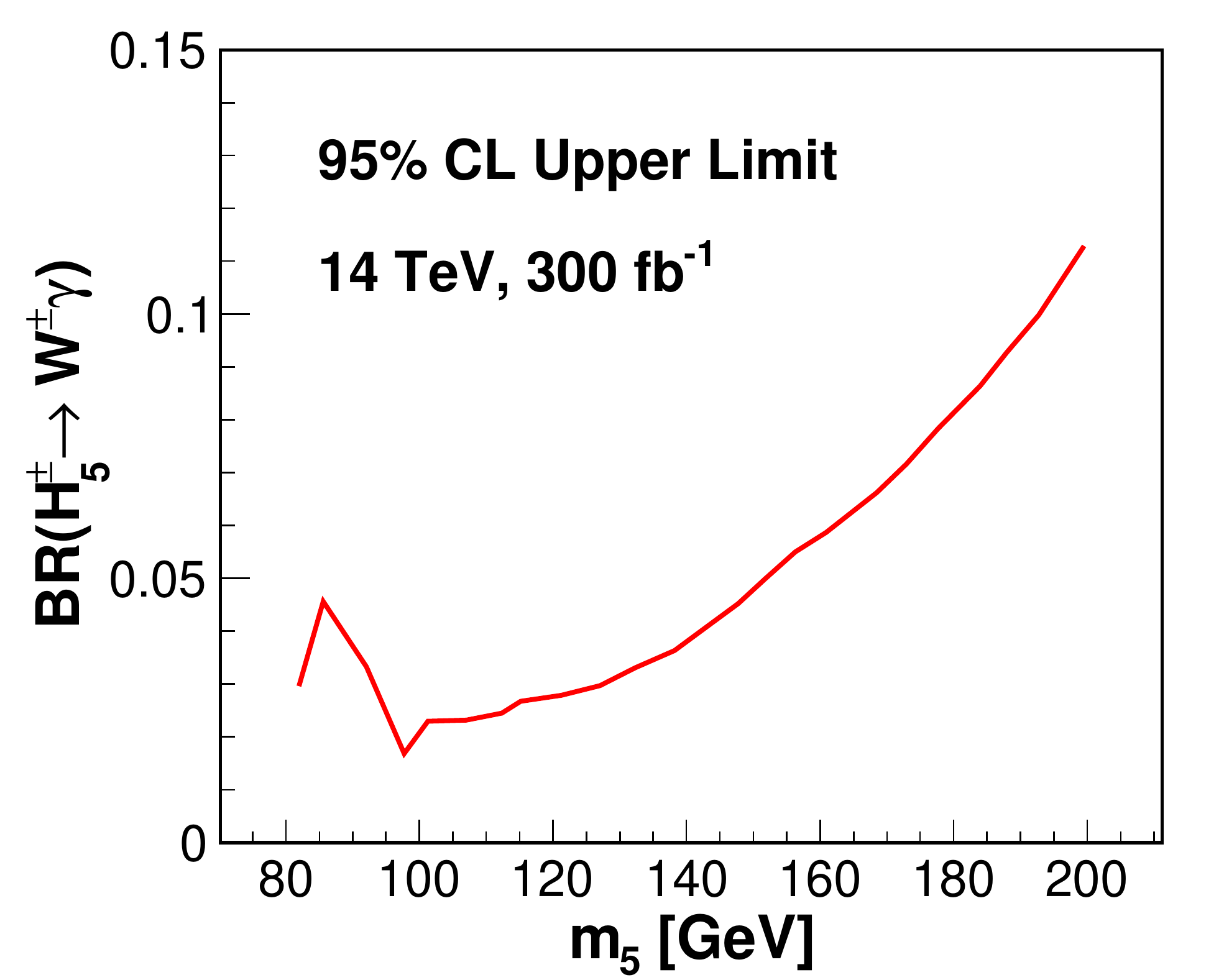}
\includegraphics[width=0.45\textwidth]{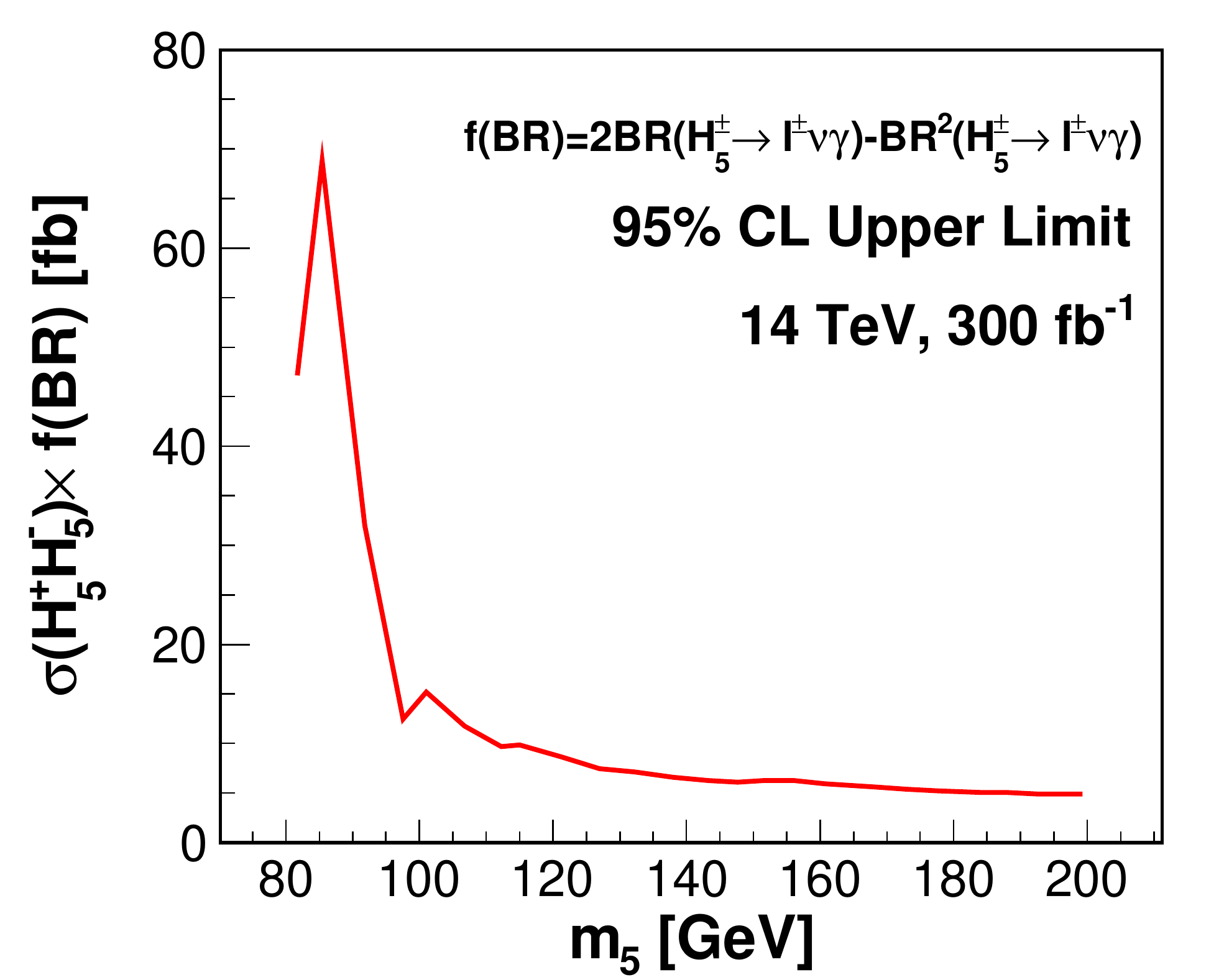}
\caption{Left: the projected 95\% CL upper limit on BR($H_5^{\pm} \to W^{\pm} \gamma$) in the GM model from the $W\gamma$ search.  Right: the projected 95\% CL upper limit on $\sigma(pp \to H_5^+ H_5^-) \times (2{\rm BR}(H_5^{\pm} \to \ell^{\pm}\nu \gamma)-{\rm BR}^2(H_5^{\pm} \to \ell^{\pm}\nu \gamma))$, assuming that $pp \to H_5^+ H_5^-$ is the only signal process.  Both plots assume 300~fb$^{-1}$ of data at the 14~TeV LHC.}
\label{fig:CSUpperlimit}
\end{figure}

The Drell-Yan cross section for production of pairs of $H_5$ scalars in the GM model depends only on the mass of $H_5$.  Thus the interpretation of the LHC exclusion in this model depends only on the branching fraction of $H_5^\pm\to W^\pm\gamma$. The projected upper limit on ${\rm BR}(H_5^\pm\to W^\pm\gamma)$ is shown in the left panel of Fig.~\ref{fig:CSUpperlimit}, where the nonlinear dependence on the branching fraction of the total cross section in Eq.~(\ref{eq:fiducial}) has been taken into account.  The projected exclusion ranges from ${\rm BR}(H_5^\pm\to W^\pm\gamma)$ of about 2\% for $m_5 \sim 100$~GeV to about 12\% for $m_5 = 200$~GeV.

In the right panel of Fig.~\ref{fig:CSUpperlimit} we show the projected 95\% confidence level upper limit on the $H_5^+H_5^-$ process alone.  The $y$ axis shows the projected upper bound on $\sigma(pp \to H_5^+ H_5^-) \times [2{\rm BR}(H_5^{\pm} \to \ell^{\pm}\nu \gamma)-{\rm BR}^2(H_5^{\pm} \to \ell^{\pm}\nu \gamma)]$.  
This can be used to estimate the sensitivity of the $W^{\pm}\gamma$ search in other models, as well as in scenarios in which the $H_5^+ H_5^-$ final state is produced resonantly through the decay of a heavier scalar particle.  (We note however that the kinematic distribution from such a decay will be different than that from Drell-Yan production, resulting in different selection efficiency.)

\subsection{Constraint on the GM model parameter space}

\begin{figure}
\centering
\includegraphics[width=0.48\textwidth]{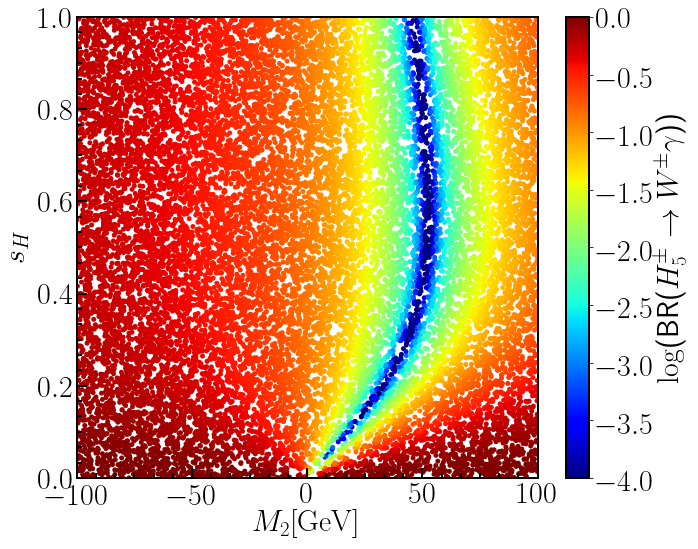}
\includegraphics[width=0.48\textwidth]{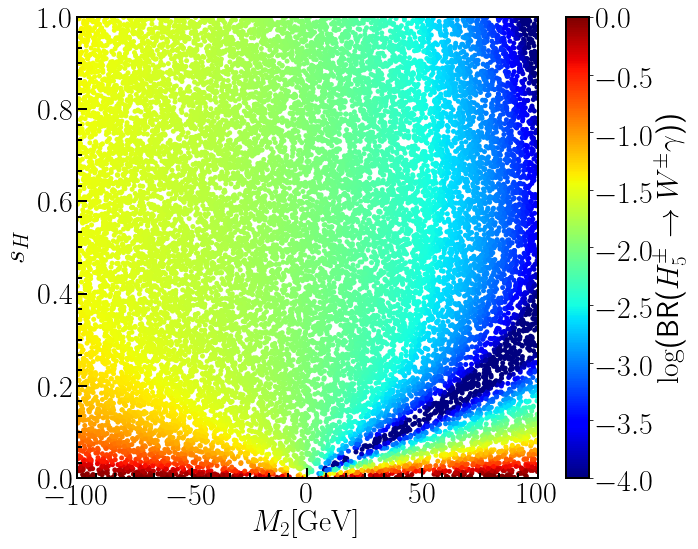}\\
\caption{Dependence of BR($H_5^\pm\to W^\pm\gamma$) on $M_2$ and $s_H$ for $m_{5} = 100$ GeV (left) and $m_{5} = 150$ GeV (right).}
\label{fig:BRinthvsM2}
\end{figure}

The projected upper bound on BR($H_5^{\pm} \to W^{\pm} \gamma$) shown in the left panel of Fig.~\ref{fig:CSUpperlimit} can be reinterpreted as a constraint on the GM model parameter space.  The dependence of BR($H_5^{\pm} \to W^{\pm} \gamma$) on the underlying parameters is remarkably simple when $m_3 \gg m_5$.  We show this as a function of $M_2$ and $s_H$ in Fig.~\ref{fig:BRinthvsM2}, for $m_5 = 100$~GeV (left) and 150~GeV (right) and the remaining model parameters chosen as in Eq.~(\ref{eq:params}).\footnote{For the sake of illustration, to populate the full range of these plots we ignore the theoretical constraints on the GM model parameters~\cite{Hartling:2014zca}.  The theoretical constraints will be satisfied in the low-$s_H$ region that we focus on below.}

For small enough $s_H \lesssim 0.3$ and fixed $m_5$, BR($H_5^{\pm} \to W^{\pm}\gamma$) depends to a good approximation only on the ratio $s_H/M_2$.  This happens because the $s_H$--suppressed terms in the triple-scalar couplings involved in $H_5^{\pm} \to W^{\pm} \gamma$ can be ignored, so that the scalar loop contribution depends only on $M_2$ as described in Appendix~\ref{app:FormFactor}.  Indeed, the most striking feature of Fig.~\ref{fig:BRinthvsM2} is the stripe in which BR($H_5^{\pm} \to W^{\pm} \gamma$) is heavily suppressed -- this is due to a cancellation between the scalar loop and the gauge and mixed gauge/scalar loop contributions to the amplitude for $H_5^{\pm} \to W^{\pm} \gamma$.  The cancellation happens only for positive $M_2$ when $s_H/M_2 \sim 10^{-2}/{\rm GeV}$ for $m_5= 100$~GeV.  The other feature of Fig.~\ref{fig:BRinthvsM2} is the $m_5$ dependence: as expected, BR($H_5^{\pm} \to W^{\pm}\gamma$) is largest when $m_5$ is well below the $WZ$ threshold; nearer the threshold, this decay only dominates when $s_H \ll 1$, and the cancellation between scalar and gauge amplitudes happens at a smaller $s_H/M_2$ value for larger mass.

\begin{figure}
\centering
\includegraphics[width=0.6\textwidth]{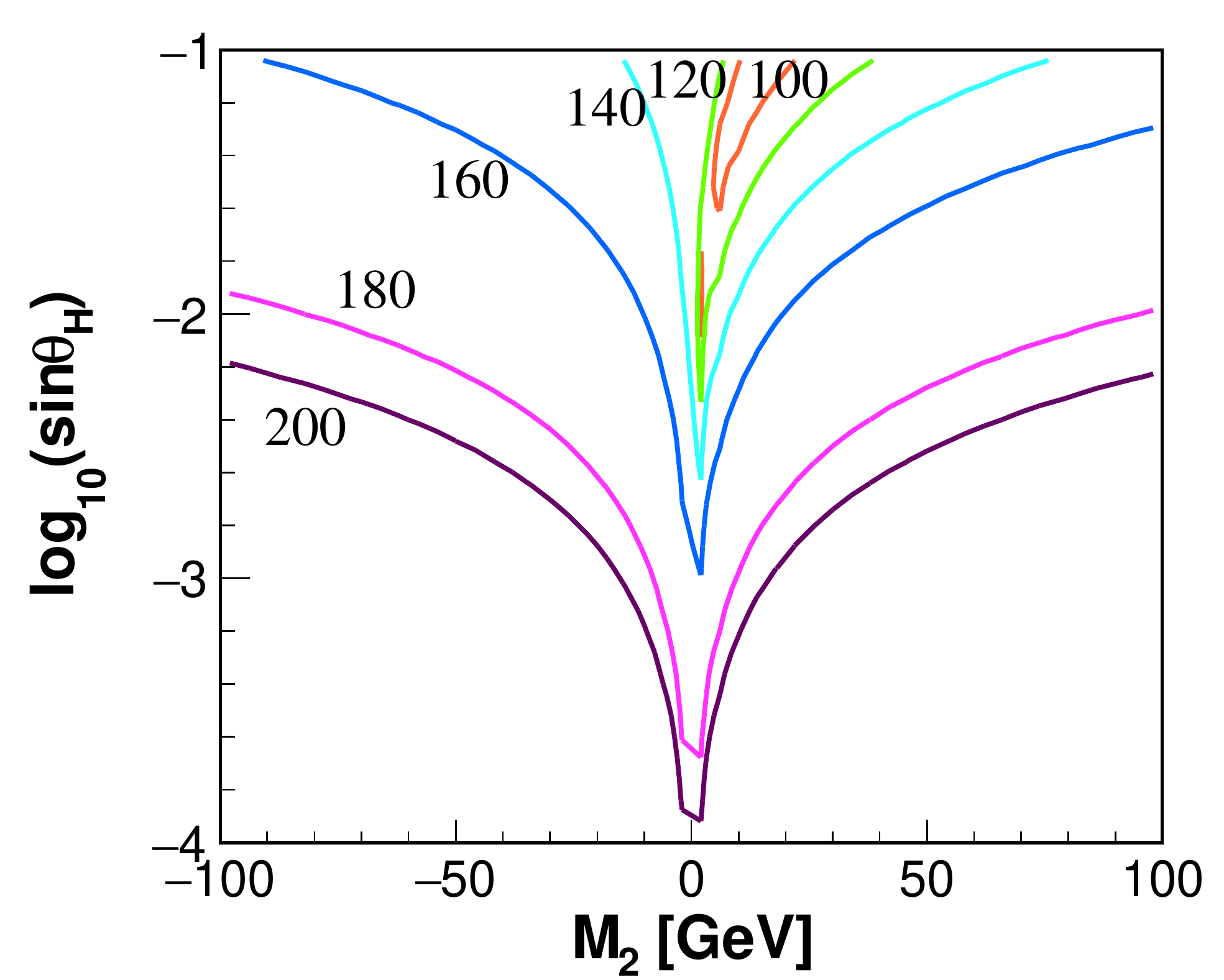}
\caption{The projected 95\% CL exclusion reach for the $W\gamma$ channel for various values of $m_{5}$ (in GeV), as a function of $M_2$ and $s_H$. The region below each line can be excluded with 300~fb$^{-1}$ of data at the 14~TeV LHC.}
\label{fig:exclu_thM2}
\end{figure}

We translate this into a projected exclusion reach in the GM model parameter space in two ways.  First, in Fig.~\ref{fig:exclu_thM2} we show the excluded region in the $M_2$--$s_H$ plane for $m_5$ values between 100 and 200~GeV in steps of 20~GeV.  The region \emph{below} each contour can be excluded by the $W\gamma$ search.  Note in particular that the $W \gamma$ channel is most sensitive at low $s_H$; this is in contrast to searches for $H_5$ produced in vector boson fusion, which lose sensitivity at low $s_H$ because the vector boson fusion cross section is proportional to $s_H^2$.

\begin{figure}
\centering
\includegraphics[width=0.45\textwidth]{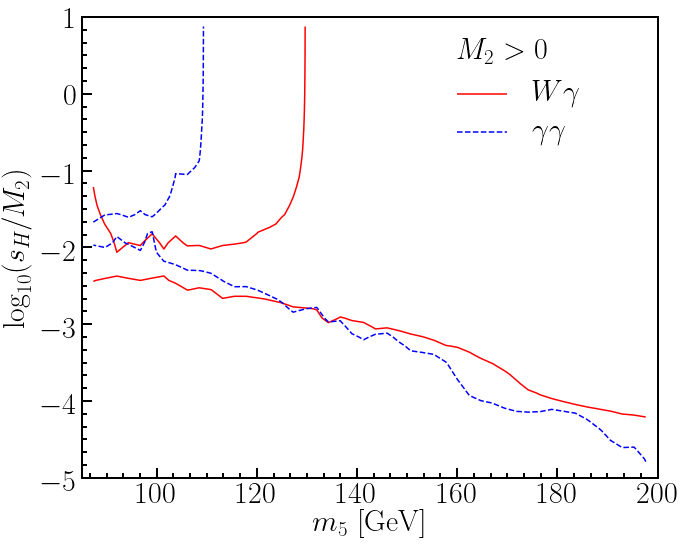}
\includegraphics[width=0.45\textwidth]{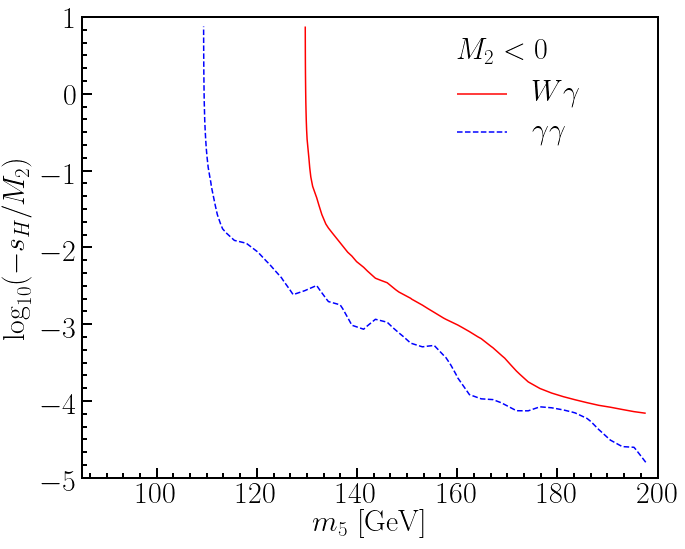}
\caption{The projected 95\% CL exclusion reach for the $W \gamma$ channel (red solid line) valid for $s_H \lesssim 0.3$ and $m_5 > 100$~GeV.  The region to the left of the curve can be excluded with 300~fb$^{-1}$ of data at the 14~TeV LHC.  Contours are shown as a function of $m_5$ and $s_H/M_2$ (in GeV$^{-1}$) for positive (left) and negative (right) $M_2$.  The region to the left of the blue dashed line is already excluded in the GM model by LHC diphoton resonance searches via the process $pp \to H_5^{\pm} H_5^0$ with $H_5^0 \to \gamma\gamma$ (see Sec.~\ref{sec:gaga}).}
\label{fig:exclu_thoverM2}
\end{figure}

Second, for small $s_H \lesssim 0.3$, we can take advantage of the fact that BR($H_5^{\pm} \to W^{\pm}\gamma$) depends to a good approximation only on the ratio $s_H/M_2$ and plot a projected exclusion in the $m_5$--$s_H/M_2$ plane.  This is shown by the red curves in Fig.~\ref{fig:exclu_thoverM2} for positive and negative $M_2$ values.  The region to the left of the curves can be excluded, except for $m_5$ values below 100~GeV where our analysis becomes unreliable.  Note the narrow unexcluded region at low $m_5$ for positive $M_2$ and $s_H/M_2 \sim 10^{-2}/{\rm GeV}$: this corresponds to the cancellation between the scalar and gauge amplitudes in $H_5^{\pm} \to W^{\pm}\gamma$ that appears as the stripe in Fig.~\ref{fig:BRinthvsM2}.  Except for this narrow region, the $W\gamma$ channel will be able to exclude $m_5$ below about 130~GeV for almost any values of $s_H/M_2$, and masses up to 200~GeV (and beyond) for sufficiently small values of $s_H/M_2$.  

\subsection{Competing constraints}
\label{sec:gaga}

There are competing constraints on the GM model for $m_5 < 200$~GeV arising from other diboson searches.  The most important of these are:
\begin{itemize}
\item[$(i)$] an 8~TeV ATLAS measurement of the $W^{\pm}W^{\pm}$ cross section in vector boson fusion~\cite{Aad:2014zda}, which was recast in Ref.~\cite{Chiang:2014bia} as a constraint on $H_5^{\pm\pm}$ production, excluding a parameter region with $s_H \gtrsim 0.4$ for $m_5 \gtrsim 140$~GeV;
\item[$(ii)$] a LEP search for $e^+e^- \to Z H$ with fermiophobic $H \to \gamma\gamma$~\cite{ALEPH:2002gcw}, which was interpreted as a constraint on $H_5^0$ in the GM model in Ref.~\cite{Degrande:2017naf}, excluding most of the parameter region with $s_H \gtrsim 0.1$ for $m_5 \lesssim 110$~GeV;
\item[$(iii)$] 8~TeV ATLAS~\cite{Aad:2014ioa} and CMS~\cite{Khachatryan:2015qba} searches for scalar diphoton resonances in the mass range 65--600~GeV and 150--850~GeV respectively.  The ATLAS search~\cite{Aad:2014ioa} quotes an upper limit on the fiducial cross section, which can be applied to Drell-Yan production of $H_5^0$ to constrain arbitrarily small values of $s_H$ in the GM model at low $m_5$, as was first pointed out in Ref.~\cite{Delgado:2016arn}.  
\end{itemize}

Searches $(i)$ and $(ii)$ put upper bounds on $s_H$ and are complementary to the $W\gamma$ search that we consider here.  Search $(iii)$ on the other hand, which relies on the loop-induced $H_5^0 \to \gamma\gamma$ channel, already directly constrains the parameter region of interest for the $W \gamma$ search.  The direct comparability of the Drell-Yan $H_5^{\pm} \to W^{\pm} \gamma$ and Drell-Yan $H_5^0 \to \gamma\gamma$ channels depends critically on the mass degeneracy of $H_5^{\pm}$ and $H_5^0$, which is a consequence of the custodial symmetry in the GM model, but need not hold in other models with fermiophobic charged Higgs bosons.

\begin{figure}
\centering
\includegraphics[width=0.48\textwidth]{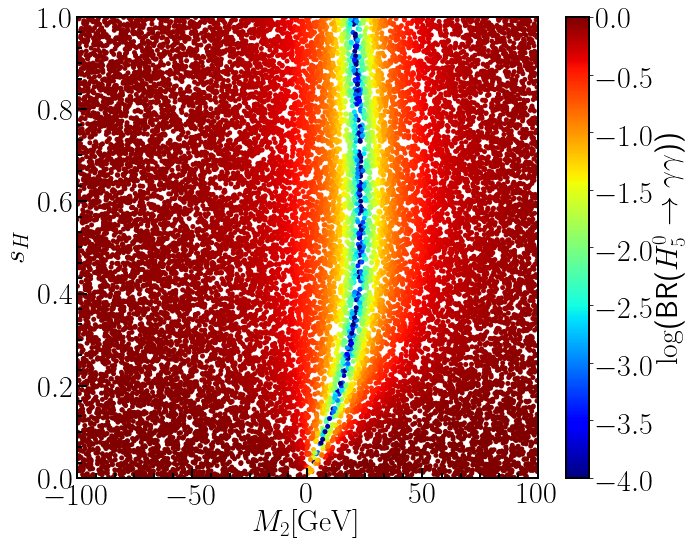}
\includegraphics[width=0.48\textwidth]{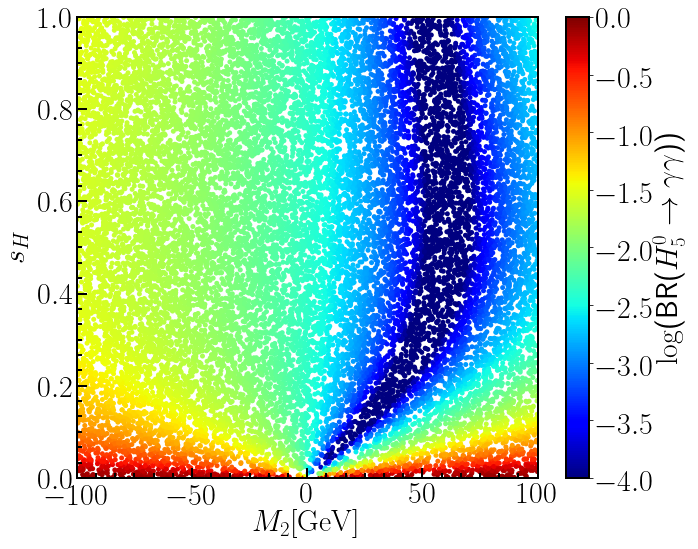}\\
\caption{Dependence of BR($H_5^0\to\gamma\gamma$) on $M_2$ and $s_H$ for $m_{5} = 100$ GeV (left) and $m_{5} = 150$ GeV (right).}
\label{fig:BRgavsM2}
\end{figure}

The branching ratio for $H_5^0 \to \gamma\gamma$ is shown in Fig.~\ref{fig:BRgavsM2} as a function of $M_2$ and $s_H$, for $m_5 = 100$~GeV (left) and 150~GeV (right).  These plots look very similar to the corresponding plots for BR($H_5^{\pm} \to W^{\pm}\gamma$) in Fig.~\ref{fig:BRinthvsM2} because the physics is mostly the same: the loop-induced $H_5^0$ decay to $\gamma\gamma$ competes with tree-level decays to $W^+W^-$ and $ZZ$ with partial widths proportional to $s_H^2$, and the decay to $\gamma\gamma$ is induced by loops of charged scalars (with an amplitude proportional to $M_2$ for $s_H$ sufficiently small) and $W$ bosons (with an amplitude proportional to $s_H$).  The cancellation between the scalar and gauge loop diagrams happens at a slightly different place in parameter space than for $H_5^{\pm} \to W^{\pm} \gamma$.  The $\gamma\gamma$ branching fraction is largest when $m_5$ is well below the $WW$ threshold; nearer the threshold, this decay only dominates when $s_H \ll 1$.

We translate the diphoton resonance search limit in Ref.~\cite{Aad:2014ioa} into a constraint on our parameter space using our simulated events for $pp \to H_5^{\pm} H_5^0$, with $H_5^0 \to \gamma\gamma$.  We decay $H_5^{\pm}$ to $W^{\pm} \gamma$ as before; in this case there is no combinatoric background to worry about because the search in Ref.~\cite{Aad:2014ioa} considered all pairs of photons for each mass hypothesis.  We obtain an efficiency as a function of $m_5$ by applying the selection from Ref.~\cite{Aad:2014ioa}: two photons with $E_T>22$~GeV and $|\eta|<2.37$ are required; if $m_{\gamma\gamma}>110$~GeV, the additional selections $E_T^{\gamma_1}/m_{\gamma\gamma}>0.4$ and $E_T^{\gamma_2}/m_{\gamma\gamma}>0.3$ are also imposed. 
We then translate the upper bound on $\sigma(pp \to H_5^{\pm} H_5^0) \times {\rm BR}(H_5^0 \to \gamma\gamma)$ into a bound in the plane of $m_5$ and $s_H/M_2$, valid for $s_H \lesssim 0.3$.  This is shown as the blue dashed line in Fig.~\ref{fig:exclu_thoverM2}; the region to the left of the line is excluded.  At large values of $s_H/M_2$, the $W$ loop contribution to $H_5^0 \to \gamma\gamma$ dominates, and the current LHC diphoton resonance searches exclude $m_5 < 110$~GeV, as pointed out already in Refs.~\cite{Delgado:2016arn,Vega:2018ddp}.  For positive $M_2$ and $s_H/M_2 \sim 10^{-2}$~GeV$^{-1}$, the scalar and gauge loops interfere destructively, resulting in a gap in the exclusion.  For smaller values of $s_H/M_2$, the scalar loop contributions dominate and the excluded region expands to higher $m_5$ as $s_H/M_2$ decreases.

We conclude that the projected exclusion reach of the $W\gamma$ channel with 300~fb$^{-1}$ at the 14~TeV LHC extends to charged Higgs masses substantially beyond the current diphoton exclusion for most values of $s_H/M_2$, except in the region in which the cancellation between the scalar and gauge amplitudes suppresses the amplitude for $H_5^+ \to W^+ \gamma$.
The two searches are complementary in two ways.  First, the cancellation in the $H_5^0 \to \gamma\gamma$ decay width happens at a slightly higher value of $s_H/M_2$ than that in $H_5^{\pm} \to W^{\pm} \gamma$, so that the $W\gamma$ channel can be used to partially close the gap in the $\gamma\gamma$ exclusion due to this destructive interference.  Second, the exclusion from $H_5^0 \to \gamma\gamma$ holds reliably for $m_5 < 100$~GeV, while our $W\gamma$ result should not be trusted in this mass range.

\section{Conclusions}
\label{sec:conclusions}

In this paper we studied the prospects for charged Higgs boson searches in the $W \gamma$ decay channel.  This loop-induced decay channel can be important if the charged Higgs is fermiophobic, particularly when its mass is below the $WZ$ threshold.  We identify useful kinematic observables and evaluate the future LHC sensitivity to this channel using the custodial-fiveplet charged Higgs in the GM model as a fermiophobic benchmark.  

We showed that the LHC with 300~fb$^{-1}$ of data at 14~TeV should be able to exclude charged Higgs masses below about 130~GeV for almost any value of $s_H$, and masses up to 200~GeV and beyond when $s_H$ is very small.  Part of this region is already excluded by LHC searches for diphoton resonances, which are relevant because $H_5^{\pm}$ and $H_5^0$ have the same mass in the GM model.  As a byproduct, we identified the most important model parameters that control the behavior of the $W\gamma$ channel and established a benchmark that captures them.

For this analysis we created a UFO model file for the GM model including effective couplings for the loop-induced scalar decays into gauge boson pairs that are absent at tree level.  We adapted GMCALC to output the existing one-loop calculations for the effective couplings in a form that can be used with the UFO model in MadGraph5.  These tools have been made publicly available as GMCALC 1.4.0.

\begin{acknowledgments}
We thank Brigitte Vachon, Kays Haddad, Howard Haber, Pedro Ferreira, and the members of the LHC Higgs Cross Section Working Group for stimulating conversations.
This work was supported by the Natural Sciences and Engineering Research Council of Canada (NSERC).  H.E.L.\ was also supported by the grant H2020-MSCA-RISE-2014 No.\ 645722 (NonMinimalHiggs).
\end{acknowledgments}

\begin{appendix}

\section{Developing a low-$s_H$ benchmark}
\label{app:LSH}

The scalar potential for the GM model given in Eq.~(\ref{equ:potential}) contains 9 parameters:
\begin{eqnarray}
	\mu_2^2,\ \mu_3^2,\ \lambda_1,\ \lambda_2,\ \lambda_3,\ \lambda_4,\ \lambda_5,\ M_1,\ M_2.
\end{eqnarray}
For our study, it is more convenient to use physical masses and couplings as input parameters as much as possible.  Therefore, we would like to use the following as inputs:
\begin{eqnarray}
	v,\ s_H,\ \sin\alpha,\ m_h,\ m_H,\ m_{3},\ m_{5},\ M_1,\ M_2.
\end{eqnarray}
Here $v = (\sqrt{2}G_F)^{-1/2}$ and $m_h = 125$~GeV are fixed by experiment, while the rest can vary.
The translation between these two parameter sets can easily be obtained by inverting the fomulas for the masses in Sec.~\ref{sec:GMdef} together with the definitions $v_{\chi} = v s_H/ \sqrt{8}$, $v_{\phi} = v c_H$:
\begin{subequations}
\label{equ:masstolam}
\begin{eqnarray}
	\mu_2^2 &=& \frac{3\sqrt{2}s_{H}c_{H} M_1 v - 8c_{H} \mathcal{M}_{11}^2-2\sqrt{6}s_{H}\mathcal{M}_{12}^2}{16c_{H}}, \\
	\mu_3^2 &=& \frac{3\sqrt{2}c^2_{H}M_1 v + 9\sqrt{2}s^2_{H}M_2 v-4\sqrt{6}c_{H}\mathcal{M}_{12}^2 - 6s_{H}\mathcal{M}_{22}^2}{12s_{H}}, \\
	\lambda_1 &=& \frac{\mathcal{M}_{11}^2}{8v^2c^2_{H}}, \\
	\lambda_2 &=& \frac{-3 c_{H}(\sqrt{2}M_1 v - 4 m_{3}^2s_{H}) + 2\sqrt{6}\mathcal{M}_{12}^2}{12v^2s_{H}c_{H}}, \\
	\lambda_3 &=& \frac{c^2_{H}(\sqrt{2}M_1 v - 3 m_{3}^2s_{H}) - s_{H}(3\sqrt{2} M_2 v s_{H} - m_{5}^2)}{v^2s^3_{H}}, \\
	\lambda_4 &=& \frac{-3c^2_{H}(\sqrt{2}M_1 v -2 m_{3}^2s_{H}) + s_{H}(9\sqrt{2} M_2 vs_{H} -2 m_{5}^2)+2s_{H}\mathcal{M}_{22}^2}{6 v^2s^3_{H}}, \\
	\lambda_5 &=& \frac{2 m_{3}^2s_{H}-\sqrt{2}M_1 v}{v^2 s_{H}}, 
\end{eqnarray}
\end{subequations}
where the mass matrix for $h$ and $H$ is
\begin{eqnarray}
	\mathcal{M}^2=\left(\begin{matrix}
	\mathcal{M}_{11}^2& \mathcal{M}_{12}^2 \\
	\mathcal{M}_{12}^2 & \mathcal{M}_{22}^2
	\end{matrix}\right) = \left(\begin{matrix}
	c^2_\alpha m_h^2+s^2_\alpha m_H^2 & s_\alpha c_\alpha(m_H^2-m_h^2)\\
	s_\alpha c_\alpha(m_H^2-m_h^2) & s^2_\alpha m_h^2 + c^2_\alpha m_H^2 
	\end{matrix}\right).
\end{eqnarray}

In our study of the $H_5^{\pm} \to W^{\pm} \gamma$ decay, we focus on the parameter region with $80~{\rm GeV} < m_{5} < 200$~GeV and small $s_H$.  However, using physical parameters as input, some of the underlying Lagrangian parameters given in Eq.~(\ref{equ:masstolam}) will blow up in the limit $s_H \to 0$ unless there are some relations between the physical input parameters.  To understand this better, it is useful to express Eq.~(\ref{equ:masstolam}) as an expansion in powers of $s_H$ and keep only the terms that have negative or zero powers of $s_H$:
\begin{subequations}
\begin{align}
	\mu_2^2 &\sim -\frac{s_\alpha^2 m_H^2 + c_\alpha^2 m_h^2}{2}, \\
	\mu_3^2 &\sim \frac{\sqrt{3}M_1 v + 2s_{2\alpha}(m_h^2-m_H^2)}{2\sqrt{6}s_H} - \frac{s_\alpha^2 m_h^2 + c_\alpha^2 m_H^2}{2}, \\
	\lambda_1 &\sim \frac{m_h^2+m_H^2+c_{2\alpha}(m_h^2-m_H^2)}{16v^2}, \\
	\lambda_2 &\sim \frac{\sqrt{6}s_{2\alpha}(m_H^2-m_h^2)-3\sqrt{2}M_1v}{12v^2 s_H} + \frac{m_{3}^2}{v^2}, \\
	\lambda_3 &\sim \frac{\sqrt{2}M_1}{vs_H^3}+\frac{m_{5}^2-3m_{3}^2}{v^2s_H^2} - \frac{2M_1+6M_2}{\sqrt{2}vs_H} + \frac{3m_{3}^2}{v^2}, \\
	\lambda_4 &\sim -\frac{M_1}{\sqrt{2}vs_H^3} + \frac{s_\alpha^2m_h^2+c_\alpha^2m_H^2+3m_{3}^2-m_{5}^2}{3v^2s_H^2} + \frac{M_1 + 3 M_2}{\sqrt{2}vs_H}, \\
	\lambda_5 &\sim -\frac{\sqrt{2}M_1}{vs_H} + \frac{2m_{3}^2}{v^2}.
\end{align}
\end{subequations}

To avoid severe constraints from perturbativity of the $\lambda_i$ in the limit $s_H \to 0$, we must choose relations among the input parameters so that all possible poles in $s_H$ are cancelled. Thus, at least the following relations should be fulfilled, where $\kappa_{\alpha}$, $\kappa_H$, and $\kappa_{\lambda_3}$ are parameters of order one:
\begin{subequations}
\begin{align}
	s_\alpha &= \kappa_\alpha s_H, \\
	m_H^2 &= \frac{3m_{3}^2-m_{5}^2}{2} + \kappa_H v^2 s_H^2, \\
	M_1 &= \frac{3 m_3^2 - m_5^2}{\sqrt{2} v}s_H + 3M_2s_H^2 + \kappa_{\lambda_3}vs_H^3.
\end{align}
\end{subequations}
Based on scans over the full set of parameters, we adopt the values
\begin{align}
\kappa_\alpha &= -0.15 - \frac{m_5}{1000 \text{ GeV}}, \nonumber \\
\kappa_H &= - \frac{m_5}{100 \text{ GeV}}, \nonumber \\
\kappa_{\lambda_3} &= -\frac{\kappa_H^2}{10}.
\end{align}
Varying these parameters has essentially no effect on the $H_5^{\pm} \to W^{\pm} \gamma$ phenomenology.

This leaves only four physical input parameters, which can be chosen as follows: two parameters $m_5$ and $\delta m^2$ that control the mass spectrum of the heavy Higgs bosons, and two parameters $s_H$ and $M_2$ that control the decays of $H_5^{\pm}$ into the competing $W^{\pm} \gamma$ and $W^{\pm} Z$ channels.  In particular, we define our benchmark as
\begin{eqnarray}
	m_5 &\in& [80,200] \text{ GeV}, \nonumber\\
	\delta m^2 &=& (300 \text{ GeV})^2, \nonumber\\
	M_2 &\in& [-100,100] \text{ GeV}, \nonumber\\
	s_H &\ll& 1, \nonumber\\
\end{eqnarray}
and
\begin{eqnarray}
	m_3^2 &=& m_5^2 + \delta m^2, \nonumber \\
	m_H^2 &=& m_5^2 + \frac{3}{2}\delta m^2 + \kappa_Hv^2s_H^2, \nonumber \\
	M_1 &=& \left[ \frac{\sqrt{2}}{v}\left(m_5^2+\frac{3}{2}\delta m^2\right) + 3M_2 s_H + \kappa_{\lambda_3}vs_H^2\right] s_H, \nonumber \\
	s_\alpha &=& \kappa_\alpha s_H.
\end{eqnarray}
Our choice of $\delta m^2 = (300~{\rm GeV})^2$ puts the $H_3$ and $H$ masses well above the $H_5$ mass, allowing us to (conservatively) ignore associated production of $H_5 H_3$.  This choice also ensures that the contribution to loop-induced decays from $H_3$ in the loop is small.

\section{Decays of $H_5^\pm\to W^\pm\gamma$, $H_5^0\to\gamma\gamma$, and $H_5^0 \to Z\gamma$ for small $s_H$}
\label{app:FormFactor}

In this section we show that, in the limit $s_H \to 0$, the expressions for the one-loop decay amplitudes for $H_5^\pm\to W^\pm\gamma$, $H_5^0\to\gamma\gamma$, and $H_5^0 \to Z\gamma$ simplify greatly, and are controlled only by the coupling parameter $M_2$ along with the masses $m_5$ and $m_3$.

The complete expressions for the one-loop effective vertices for $H_5^\pm\to W^\pm\gamma$, $H_5^0\to\gamma\gamma$, and $H_5^0 \to Z\gamma$ involve loops of gauge bosons, scalars, and combinations thereof and have been computed in Ref.~\cite{Degrande:2017naf}.  The expressions for these amplitudes can be greatly simplified in the limit $s_H \to 0$, because all amplitudes involving gauge bosons in the loop vanish in this limit.  We are left with~\cite{Degrande:2017naf}
\begin{eqnarray}
	S_{H_5^\pm\to W^\pm\gamma} &\approx& \sum_{s_1,s_2}A^{H_5^+W\gamma}_{s_1s_2s_2}, \nonumber\\
	S_{H_5^0\to\gamma\gamma} &\approx& \frac{\alpha_{em}}{2\pi v}\sum_{s}\frac{C_{H_5^0ss^*}v}{2m_s^2}Q_s^2F_0(\tau_s), \nonumber \\
	S_{H_5^0\to Z\gamma} &\approx& -\frac{\alpha_{em}}{2\pi v}\sum_{s}\beta_s^{H_5^0}A_{s}^{H_5^0Z\gamma},
	\label{eq:Sapprox}
\end{eqnarray}
where the sums run over the scalars that can appear in the loop and 
\begin{eqnarray}
	A^{H_5^+W\gamma}_{s_1s_2s_2} &=& -\frac{\alpha_{em}}{\pi}\frac{C_{H_5^+s_1^*s_2}C_{W^-s_1s_2^*}Q_{s_2}}{4m_s^2}I_1(\tau_s,\bar\lambda_s), \nonumber \\
	A^{H_5^0Z\gamma}_{s} &=& 2 C_{Zss^*}Q_s I_1(\tau_s,\bar\lambda_s), \nonumber \\
	\beta_s^{H_5^0} &=& \frac{C_{H_5^0ss^*}v}{2m_s^2}.
	\label{eq:Afactors}
\end{eqnarray}
Here $\alpha_{em}$ is the electromagnetic fine structure constant and $Q_s$ is the electric charge of scalar $s$ in units of $e$.  The functions $F_0(\tau)$ and $I_1(\tau,\bar\lambda)$ are the usual scalar loop form factors that appear in Higgs decays to $\gamma\gamma$ and $Z\gamma$~\cite{Gunion:1989we},\footnote{We put a bar over the $\lambda$ in $I_1(\tau,\bar\lambda)$ to avoid confusion with the scalar quartic couplings.}
\begin{eqnarray}
	F_0(\tau_s) &=& \tau_s[1-\tau_sf(\tau_s)], \nonumber \\
	I_1(a,b) &=& \frac{ab}{2(a-b)}+\frac{a^2b^2}{2(a-b)^2}[f(a)-f(b)]+\frac{a^2b}{(a-b)^2}[g(a)-g(b)],\end{eqnarray}
where for decays of $H_5$, the arguments are
\begin{eqnarray}
	\tau_s = \frac{4m_s^2}{m_{5}^2},\qquad \bar\lambda_s = \frac{4m_s^2}{m_V^2},
\end{eqnarray}
where $V = W$ or $Z$ is the massive final-state gauge boson.  The functions $f$ and $g$ are defined in the usual way as~\cite{Gunion:1989we},
\begin{eqnarray}
f(\tau) &=& \begin{cases}
\left[\sin^{-1}\left(\sqrt{\frac{1}{\tau}}\right)\right]^2 & \text{if }\tau \geq 1, \\
-\frac{1}{4}\left[\log\left(\frac{1+\sqrt{1-\tau}}{1-\sqrt{1-\tau}}\right)-i\pi\right]^2 & \text{if }\tau < 1,
\end{cases} \\
g(\tau) &=& \begin{cases}
\sqrt{\tau-1}\sin^{-1}\left(\sqrt{\frac{1}{\tau}}\right) & \text{if }\tau\geq1, \\
\frac{1}{2}\sqrt{1-\tau}\left[\log\left(\frac{1+\sqrt{1-\tau}}{1-\sqrt{1-\tau}}\right)-i\pi\right] & \text{if } \tau<1.
\end{cases}
\end{eqnarray}

The couplings that appear in Eqs.~(\ref{eq:Sapprox}) and (\ref{eq:Afactors}) also simplify in the $s_H \to 0$ limit and are given in this limit in Table~\ref{tab:SH5WGA}.  These couplings are defined in terms of the triple-scalar and vector-scalar-scalar Feynman rules by $-iC_{H_is_1^*s_2}$ and $ieC_{Vs_1s_2}(p_1-p_2)^\mu$, respectively, with all particles incoming.

\begin{table}
\centering
\begin{tabular}{|ccc|ccc|}
\hline \hline
\multicolumn{3}{|c|}{$H_5^\pm\to W^\pm\gamma$} & \multicolumn{3}{c|}{$H_5^0\to\gamma\gamma/Z\gamma$} \\
\hline\hline
$s_1$ and $s_2$ & $C_{H_5^+s_1^*s_2}$ & $C_{W^-s_1s_2^*}$ & $s$ & $C_{H_5^0ss^*}$ & $C_{Zss^*}$\\
\hline\hline
$H_3^0$, $H_3^-$ &  $-i3\sqrt{2}M_2$  & $- i/2s_W$ & $H_3^+$ & $\sqrt{6}M_2$ & $(1-2s_W^2)/2s_Wc_W$ \\
\hline
$H_5^0$, $H_5^-$ &  $-\sqrt{6}M_2$ & $\sqrt{3}/2s_W$ & $H_5^+$ & $-\sqrt{6}M_2$ & $(1-2s_W^2)/2s_Wc_W$ \\
\hline
$H_5^-$, $H_5^{--}$ & $6M_2$  & $-1/\sqrt{2}s_W$ & $H_5^{++}$ & $2\sqrt{6}M_2$ & $(1-2s_W^2)/s_Wc_W$ \\
\hline
$H_5^{++}$, $H_5^+$ &  $6M_2$ & $1/\sqrt{2}s_W$ & & &  \\
\hline \hline
\end{tabular}
\caption{The scalars that contribute to the one-loop $H_5^\pm\to W^\pm\gamma$ and $H_5^0 \to \gamma\gamma, Z\gamma$ decays and the corresponding couplings in the limit $s_H \to 0$.}
\label{tab:SH5WGA}
\end{table}

Inserting the couplings from Table~\ref{tab:SH5WGA} and doing the sums, the form factors $S$ for $H_5^\pm\to W^\pm\gamma$, $H_5^0\to\gamma\gamma$, and $H_5^0 \to Z\gamma$ can be written in the limit $s_H \to 0$ in the relatively simple form,
\begin{eqnarray}
	S_{H_5^\pm\to W^\pm\gamma} &\xrightarrow{s_H\to0}& -\frac{\alpha_{em}}{2\pi}\frac{3\sqrt{2}}{4}\frac{M_2}{s_W}\left(\frac{I_1(\tau_3,\bar\lambda_3)}{m_{3}^2}+\frac{7I_1(\tau_5,\bar\lambda_5)}{m_{5}^2}\right), \nonumber \\
	S_{H_5^0\to\gamma\gamma} &\xrightarrow{s_H\to0}& \frac{\alpha_{em}}{2\pi}\frac{\sqrt{6}}{2}M_2\left(\frac{F_0(\tau_3)}{m_{3}^2}+\frac{7F_0(\tau_5)}{m_{5}^2}\right), \nonumber\\
	S_{H_5^0\to Z\gamma} &\xrightarrow{s_H\to0}& -\frac{\alpha_{em}}{2\pi}\sqrt{6}M_2\frac{1-2s_W^2}{2s_Wc_W}\left(\frac{I_1(\tau_3,\bar\lambda_3)}{m_{3}^2}+\frac{7I_1(\tau_5,\bar\lambda_5)}{m_{5}^2}\right).
\end{eqnarray}
We note in particular that for $s_H \to 0$, all of these form factors are proportional to $M_2$, and are otherwise controlled only by the masses $m_3$ and $m_5$.

\end{appendix}

\bibliographystyle{JHEP}
\bibliography{references}

\end{document}